\documentclass[12pt,preprint]{aastex}      

\usepackage[latin1]{inputenc}
\usepackage{color}

\shorttitle{Dust Distribution in HL~Tau}
\shortauthors{Carrasco-Gonz\'alez et al.}

\begin{document}

\title{The Radial Distribution of Dust Particles in the HL Tau Disk from ALMA and VLA Observations}

\correspondingauthor{Carlos Carrasco-Gonz\'alez}
\email{c.carrasco@irya.unam.mx}

\author{Carlos Carrasco-Gonz\'alez}
\affiliation{Instituto de Radioastronom\'{\i}a y Astrof\'{\i}sica (IRyA), Universidad Nacional Aut\'onoma de M\'exico (UNAM)}

\author{Anibal Sierra}
\affiliation{Instituto de Radioastronom\'{\i}a y Astrof\'{\i}sica (IRyA), Universidad Nacional Aut\'onoma de M\'exico (UNAM)}

\author{Mario Flock}
\affiliation{Max-Planck-Institut f\"ur Astronomie (MPIA)}

\author{Zhaohuan Zhu}
\affiliation{Department of Physics and Astronomy, University of Nevada, Las Vegas}

\author{Thomas Henning}
\affiliation{Max-Planck-Institut f\"ur Astronomie (MPIA)}

\author{Claire Chandler}
\affiliation{National Radio Astronomy Observatory (NRAO)}

\author{Roberto Galv\'an-Madrid}
\affiliation{Instituto de Radioastronom\'{\i}a y Astrof\'{\i}sica (IRyA), Universidad Nacional Aut\'onoma de M\'exico (UNAM)}

\author{Enrique Mac\'ias}
\affiliation{Department of Astronomy, Boston University}

\author{Guillem Anglada}
\affiliation{Instituto de Astrof\'{\i}sica de Andaluc\'{\i}a (IAA-CSIC)}

\author{Hendrik Linz}
\affiliation{Max-Planck-Institut f\"ur Astronomie (MPIA)}

\author{Mayra Osorio}
\affiliation{Instituto de Astrof\'{\i}sica de Andaluc\'{\i}a (IAA-CSIC)}

\author{Luis F. Rodr\'{\i}guez}
\affiliation{Instituto de Radioastronom\'{\i}a y Astrof\'{\i}sica (IRyA), Universidad Nacional Aut\'onoma de M\'exico (UNAM)}

\author{Leonardo Testi}
\affiliation{European Southern Observatory (ESO)}

\author{Jos\'e M. Torrelles}
\affiliation{Institut de Ci\`encies de l'Espai (CSIC) and Institut d'Estudis Espacials de Catalunya (IEEC)}

\author{Laura P\'erez}
\affiliation{Departamento de Astronom\'{\i}a, Universidad de Chile}

\author{Yao Liu}
\affiliation{Max-Planck-Institut f\"ur Extraterrestrische Physik}
\affiliation{Purple Mountain Observatory, Chinese Academy of Sciences}

\begin{abstract}
 Understanding planet formation requires to discern how dust grows in protoplanetary disks. An important parameter to measure in disks is the maximum dust grain size present. This is usually estimated through measurements of the dust opacity at different millimeter wavelengths assuming optically thin emission and dust opacity dominated by absorption. However, ALMA observations have shown that these assumptions might not be correct in the case of protoplanetary disks, leading to overestimation of particle sizes and to underestimation of the disk's mass. Here, we present an analysis of high quality ALMA and VLA images of the HL~Tau protoplanetary disk, covering a wide range of wavelengths, from 0.8~mm to 1~cm, and with a physical resolution of $\sim$7.35~au. We describe a procedure to analyze a set of millimeter images without any assumption about the optical depth of the emission, and including the effects of absorption and scattering in the dust opacity. This procedure allows us to obtain the dust temperature, the dust surface density and the maximum particle size at each radius. In the HL~Tau disk, we found that particles have already grown up to a few millimeters in size. We detect differences in the dust properties between dark and bright rings, with dark rings containing low dust density and small dust particles. Different features in the HL~Tau disk seem to have different origins. Planet-disk interactions can explain substructure at the external half of the disk, but the internal rings seem to be associated to the presence of snow lines of several molecules.
\end{abstract}

\section{Introduction}

 The first stages in the formation of terrestrial planets require the growth of micron-sized dust grains to cm-sized pebbles and then to km-sized planetesimals (for recent reviews on dust evolution and planet formation, see Testi et al. 2014, Johansen \& Lambrechts 2017, and references therein). Then, a necessary first step to understand planet formation is to discern where, how, and when the dust growth process starts in protoplanetary disks. 

 Dust particle sizes in protoplanetary disks are frequently estimated using the opacity properties of dust at millimeter wavelengths (e.g., Rodmann et al. 2006; Isella et al. 2010; Guilloteau et al. 2011; Mac\'{\i}as et al. 2019). At these wavelengths, it is usually assumed that the dust opacity is dominated by absorption and that the emission is optically thin. Under these assumptions, the spectral index of the emission can be easily related to the spectral behaviour of the absorption coefficient, which strongly depends on the properties of the dust particle distribution, and in particular, to the maximum particle size present in the disk (e.g., D'Alessio et al. 2001). In this way, multi-wavelength high angular resolution observations have allowed to detect radius-dependent dust properties for a handful of disks around T-Tauri stars. Grain sizes varying from millimeters in the outer parts of the disk up to several centimeters at the center of the disks have been found (e.g., P\'erez et al. 2012, 2015; Menu et al. 2014; Tazzari et al. 2016). Thus, it is usually accepted that cm-sized pebbles, a first step to form planetesimals, are already present at the internal parts of T-Tauri disks.
 
 The Atacama Large Millimeter/submillimeter Array (ALMA), with its unprecedent high sensitivity and high angular resolution at (sub)millimeter wavelengths, offers now the possibility to study the dust distribution in protoplanetary disks in great detail, with physical resolutions of only a few astronomical units. After several years of observing with ALMA we have learned that very often, protoplanetary disks present complex concentric substructures in the form of dark and bright rings (e.g., Andrews et al. 2018a; Long et al. 2018). Dark rings are usually interpreted as regions of low dust density or gaps, while bright rings are interpreted as regions of high dust density. The detection of these structures in the disks has important consequences for dust growth and planet formation processes. Large particles can be naturally accumulated and concentrated in dense rings (e.g., Flock et al. 2015; Ruge et al. 2016; Pinilla et al. 2017), which become excellent places for the growth of dust to larger sizes, and then for the formation of planetesimals. The ubiquitous presence of these structures also makes mandatory to study the properties of the dust with the highest angular resolution possible to distinguish differences in the dust properties between rings and gaps. The problem is that the high densities reached in some of the rings usually imply optically thick emission at wavelengths shorter than 3~mm (e.g., Pinte et al. 2015, Liu et al. 2017), making difficult to obtain dust properties directly from millimeter observations. Thus, in recent years, it has become clear that complementary high quality observations at longer wavelengths, where emission is optically thinner, are fundamental in order to study the earliest stages of the planetary formation process (e.g. Carrasco-Gonz\'alez et al. 2016; Mac\'{\i}as et al. 2017, 2018; Andrews et al. 2018b). Moreover, the usual assumption that the dust opacity is dominated only by absorption, while a reasonable approximation for the interstellar medium, might be very far from reality in the case of protoplanetary disks. We should expect dust particles with sizes of millimeters or larger in protoplanetary disks, at least in those parts where dust growth is effective, and for those large particles the dust opacity at millimeter wavelengths should be indeed dominated by scattering (e.g., D'Alessio et al. 2001; Sierra et al. 2019). These two assumptions, optically thin emission and absorption-dominated opacity, could easily lead to overestimations of the dust particle sizes in protoplanetary disks (e.g., Liu et al. 2019; Zhu et al. 2019). 
 
 Very recently, an alternative method to study the particle distribution in protoplanetary disks has been proposed and applied to several objects. The properties of the dust scattering strongly depend on the maximum grain size of the dust distribution. Thus, by observing self-scattering dust polarization at different wavelengths, it is possible to estimate the maximum dust grain size in the disk (Kataoka et al. 2015; Pohl et al. 2016). In principle, this method promises a direct measurement of the particle sizes in the disk without worrying about the optical depth of the emission. However, the expected polarization degrees due to self-scattering in protoplanetary disks are extremely low, of only a few percent of the total dust emission at millimeter wavelengths (e.g., Stephens et al. 2019), which makes very difficult to study polarization at high angular resolutions even with ALMA. Nevertheless, polarization observations with moderate angular resolutions (several hundreds of m.a.s.) have been succesful in constraining the global maximum particle size in some T-Tauri disks. Surprisingly, in most cases, while studies based on measurements of the opacity predict millimeter-sized dust particles as minimum, works based on dust polarization suggest particle sizes of only a few hundreds of microns at most (e.g., Kataoka et al. 2017; Bacciotti et al. 2018). If the polarization measurements are accurate, this would imply that the dust growth process is much slower than suggested by studies based on opacity measurements.

 Probably the most representative example of how the study of protoplanetary disks has evolved in the last years is the HL~Tau disk. Located in the Taurus-Auriga star forming region, this is one of the youngest ($\lesssim$1~Myr; van der Marel et al. 2019) T-Tauri stars observed with ALMA at the highest angular resolution. The protoplanetary disk surrounding HL~Tau is one of the first disks where multiple gaps and rings were observed in the dust distribution (ALMA Partnership et al. 2015b). Early radiative transfer modeling of the ALMA images showed that emission from the central part of the disk and from most of the rings is optically thick (Jin et al. 2016; Pinte et al. 2016). Subsequent very high sensitivity observations at longer wavelengths with the Very Large Array (VLA) showed that emission at wavelengths of 7.0~mm and longer should be optically thin in the whole disk (Carrasco-Gonz\'alez et al. 2016). These observations also revealed a radial gradient in the spectral index of the emission between the optically thinnest available wavelengths (7.0 and 3.0~mm), which strongly suggested radius-dependent dust properties (Carrasco-Gonz\'alez et al. 2016). A more recent radiative transfer modeling, now including the available ALMA and VLA images, found that the bright rings are most probably associated with higher densities and larger dust particles than the dark rings (Liu et al. 2017). A recent analysis of the dust polarization at several wavelengths imposed a maximum grain size of 100 microns for the whole disk (Kataoka et al. 2017), which was surprising since particles of at least several millimeters were expected from the values of the spectral indices between 7.0 mm and 3.0 mm.

 In this paper, we present an analysis of high quality ALMA and VLA images of the HL~Tau protoplanetary disk, covering a wide range of wavelengths, from 0.8~mm to 1~cm. This set of images includes previously reported, as well as new ALMA and VLA observations. We use these data to fit the spectral energy distribution (SED) of the millimeter emission at each radius, without any assumption about the optical depth of the emission, and including the effect of scattering in the dust opacity. We describe a procedure that allows us to simultaneously obtain at each radius of the disk, the optical depth of the emission, the albedo at millimeter wavelengths, and the spectral behaviour of the dust opacity. From these, we obtain the radial particle size distribution in the disk with a physical resolution of 7.35 au. We found that particles in the very young disk around HL~Tau have already grown up to a few millimeters in size, i.e., larger than suggested by the polarization studies ($\sim$100~$\mu$m), but smaller than typical sizes found in other T-Tauri disks (a few centimeters). Our angular resolution allows us to detect differences in the grain size distribution between dark and bright rings, with relatively smaller particles located in the dark rings. We discuss their characteristics in the context of the formation mechanism of the rings. Based on our results, we suggest that different rings could have different origins.

\section{Observations and image analysis}

\begin{figure}[p]
\begin{center}
\includegraphics[width=\textwidth]{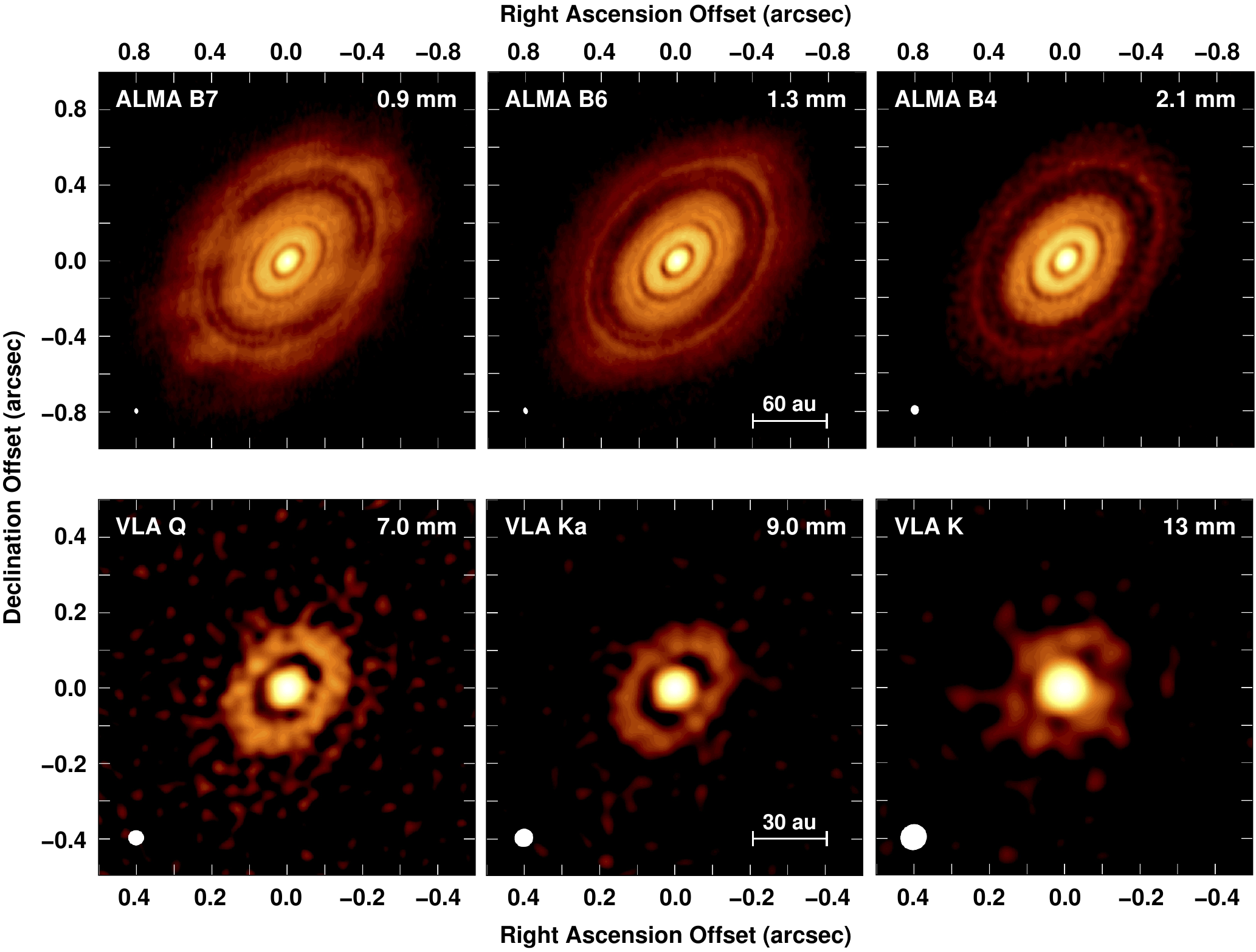}
\caption{\footnotesize{Initial set of HL~Tau images (see \S\ref{ObsALMA} and \S\ref{ObsVLA} for details). The top panels show ALMA images of the entire HL~Tau disk (2$\arcsec$$\times$2$\arcsec$), while the bottom panels show VLA images of the central part of the disk (1$\arcsec$$\times$1$\arcsec$). Color scale in all panels are from 2 times the rms noise of each map to the maximum intensity of the map (see Table \ref{Tab1}). }}
\label{Fig1}
\end{center}
\end{figure}

 We used ALMA and VLA\footnote{The National Radio Astronomy Observatory is a facility of the National Science Foundation operated under cooperative agreement by Associated Universities, Inc} data at several wavelengths described in subsections \ref{ObsALMA} and \ref{ObsVLA}. Parameters for representative images obtained from these data are given in Table \ref{Tab1} and images are shown in Figure \ref{Fig1}. Images obtained with the VLA are corrected for free-free emission from the HL~Tau radio jet as described in Appendix \ref{ObsFreeFree}. In subsection \ref{FinalImages} we discuss the final images used in the analysis. Parameters of final images used for the analysis are given in Table \ref{Tab2} and the images are shown in Figure \ref{Fig2}. Radial profiles of the brightness temperature at each wavelength, and spectral indices between the different images are shown in Figure \ref{Fig3}.

\begin{deluxetable}{cccccc}
\tablecaption{Initial ALMA and VLA images of HL Tau\label{Tab1}}
\tablehead{
\colhead{} & \colhead{} & \colhead{Wavelength} & \colhead{rms noise} & \colhead{Maximum Intensity} & \colhead{Synthesized Beam Size ; P.A.}\\
\colhead{Telescope} & \colhead{Band} & \colhead{(mm)} & \colhead{($\mu$Jy~beam$^{-1}$)} & \colhead{(mJy~beam$^{-1}$)} & \colhead{(mas x mas ; deg)}
}
\startdata
ALMA      &   B7   &     0.9      &    20   &  1.15      &    30 $\times$ 19 ;  4 \\
ALMA      &   B6   &     1.3      &    12   &  6.5       &    35 $\times$ 22 ; 11 \\
ALMA      &   B4   &     2.1      &    14   &  4.3       &    49 $\times$ 43 ;  4 \\
VLA       &    Q   &     7.0      &     4   &  0.416     &    42 $\times$ 40 ; 90 \\
VLA       &   Ka   &     9.0      &    3.5  &  0.741     &    49 $\times$ 47 ; $-$60 \\
VLA       &    K   &    13.0      &    3.1  &  0.357     &    71 $\times$ 67 ; $-$61 \\
\enddata
\tablecomments{ALMA images are the reference images from the long baselines science verification campaign. ALMA B4 image is the self calibrated one. VLA images are obtained from robust=0 weighting.}
\end{deluxetable}

\subsection{ALMA Observations} \label{ObsALMA}

 We used images obtained with ALMA at bands 7, 6, and 4, corresponding to central wavelengths of 0.9, 1.3 and 2.1 mm, respectively. Images at bands 7 and 6 were obtained in 2014 as part of the ALMA Long Baseline Science Verification program (ALMA Partnership et al. 2015a) and are the iconic images previously reported and discussed by ALMA Partnership et al. (2015b). We downloaded the reference images from the ALMA Science Verification webpage\footnote{https://almascience.nrao.edu/alma-data/science-verification}. These images have angular resolutions of $\sim$30 and $\sim$35~m.a.s., and rms noises of $\sim$20 and $\sim$12 $\mu$Jy~beam$^{-1}$, respectively (see Table \ref{Tab1}). The band~4 data was obtained in a more recent (November 2015) ALMA observation, also part of the Long Baselines Science Verification program, but it has not been reported before. We downloaded the uv data and followed instructions from ALMA staff for applying the calibration, self-calibration and imaging. The final, self-calibrated band 4 image has an angular resolution of $\sim$49~m.a.s., and an rms noise of $\sim$14~$\mu$Jy~beam$^{-1}$ (for a robust 0 weighted image).

\subsection{VLA observations} \label{ObsVLA}

 We used high sensitivity VLA observations at Q, Ka and K bands, corresponding to central wavelengths of 7.0, 9.0 and 13 mm, respectively. The Q band data is a combination of C, B and A configuration data and were already presented and discussed in Carrasco-Gonz\'alez et al. (2016). The 7.0 mm image (robust 0 weighted image) has an angular resolution of $\sim$42~m.a.s. and an rms noise of $\sim$4~$\mu$Jy/beam. We obtained new observations at Ka and K bands during 7 observational sessions between October 1st and 18th, 2016, using the A configuration of the VLA. These data were calibrated using the NRAO pipeline for VLA continuum data. In all sessions, flux, bandpass and complex gain calibrations were performed by observing 3C147, 3C84 and J0431+1731, respectively. These were the same calibrators used in the previous Q band data. Angular resolutions and rms noises for robust 0 weighted images are $\sim$49~m.a.s. and $\sim$3.5~$\mu$Jy/beam for Ka band and $\sim$70~m.a.s. and $\sim$3.1 $\mu$Jy/beam for K band. The central $\sim$15 m.a.s. in the VLA images at all bands are strongly affected by free-free emission from the well known HL~Tau radio jet. We used all the available high angular resolution data to model and subtract the jet emission in the uv-plane (see Appendix \ref{ObsFreeFree} for details on this procedure).

\subsection{Final images and radial profiles} \label{FinalImages}

\begin{deluxetable}{ccccc}
\tablecaption{Final ALMA and VLA images of HL Tau used in the analysis\label{Tab2}}
\tablehead{
\colhead{}          & \colhead{} & \colhead{Wavelength} & \colhead{rms noise} & \colhead{Maximum Intensity} \\
\colhead{Telescope} & \colhead{Band} & \colhead{(mm)} & \colhead{($\mu$Jy~beam$^{-1}$)} & \colhead{(mJy~beam$^{-1}$)}
}
\startdata
ALMA      &   B7   &     0.9      &   15  & 33         \\
ALMA      &   B6   &     1.3      &   20  & 15         \\
ALMA      &   B4   &     2.1      &   15  &  5        \\
VLA       &   Ka+Q &     8.0      &    3  &  0.2        \\
\enddata

\tablecomments{All images are convolved to a circular beam with a size of 50 m.a.s. VLA image is free-free corrected, and was obtained by combining Ka and Q bands.}

\end{deluxetable}

\begin{figure}[p]
\begin{center}
\includegraphics[width=\textwidth]{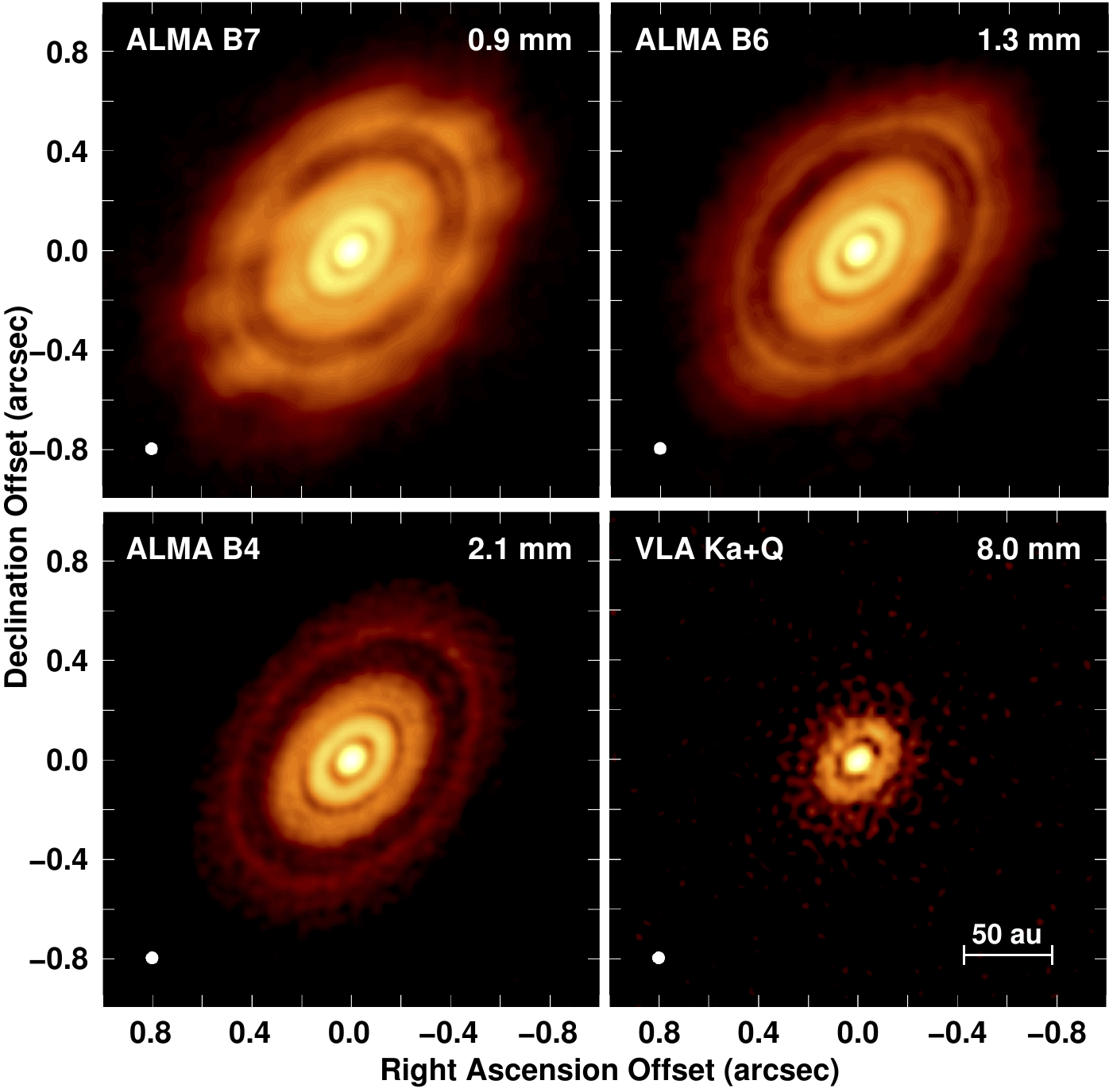}
\caption{\footnotesize{Final images of the HL~Tau disk used in the analysis (see \S\ref{FinalImages} for details). All images are convolved to the same angular resolution of 50 m.a.s. ($\sim$7.35~au), and all panels show the same spatial extension (2$\arcsec$ $\times$ 2$\arcsec$, centered on the disk peak intensity). Color scales in the four panels to the left are between 2 times the rms of each map and the maximum intensity (see Table \ref{Tab2}). The VLA image at 8.0 mm was made combining data at Ka and Q bands after subtraction of the free-free emission (see \S\ref{ObsFreeFree} for details).}}
\label{Fig2}
\end{center}
\end{figure}

\begin{figure}[p]
\begin{center}
\includegraphics[width=\textwidth]{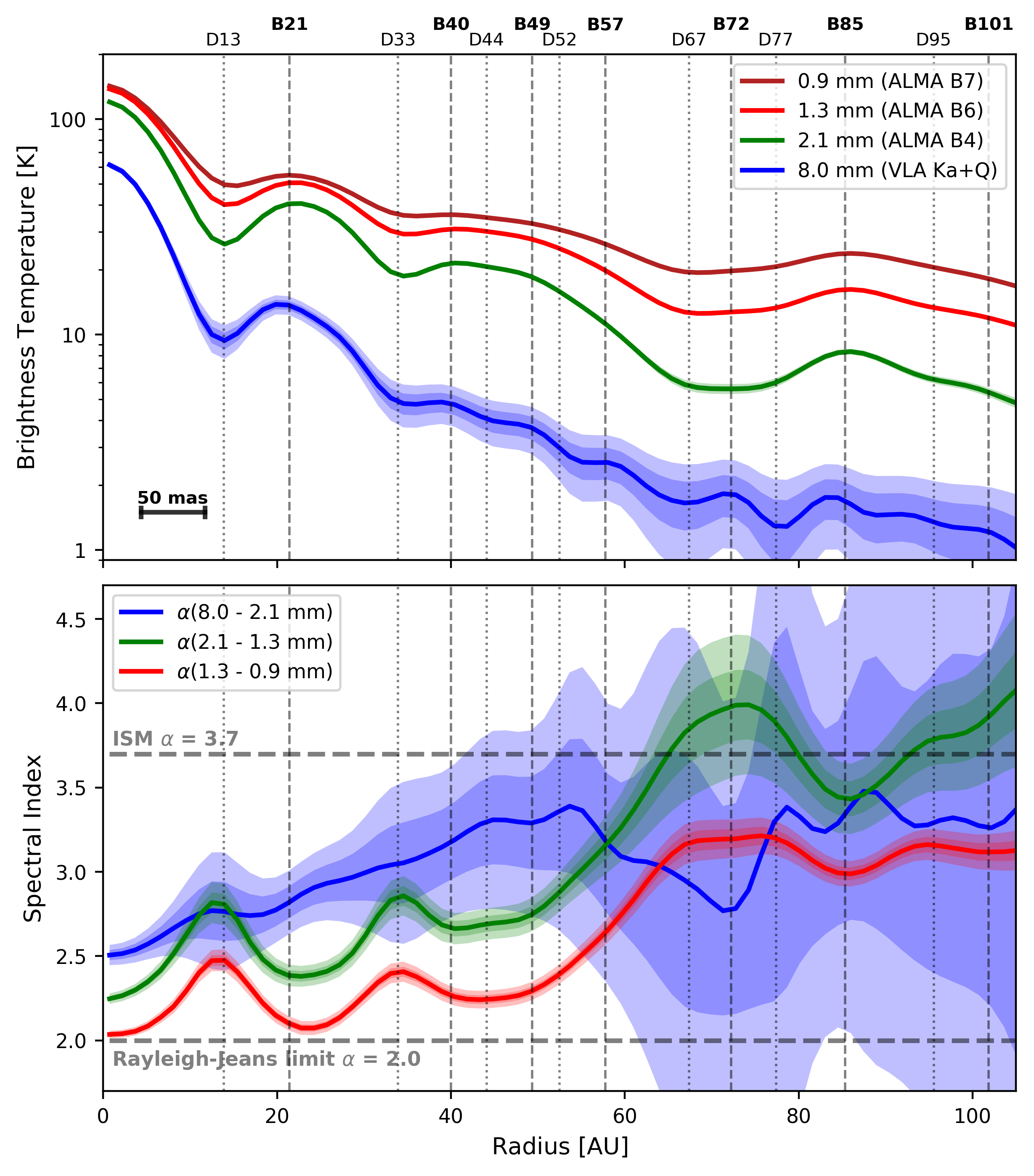}
\caption{Radial profiles of the Brightness Temperature in the ALMA and VLA images (top panel) and spectral indices between different combinations of wavelengths (bottom panel). In both panels, statistical errors (based on the rms noise of the maps) at 1- and 2-$\sigma$ levels are shown as filled bands around the nominal values.}
\label{Fig3}
\end{center}
\end{figure}

 In order to compare images at different wavelengths, it is necessary to have all of them at the same angular resolution. The weighting of the visibilities affects the sensitivity and angular resolution of the images, with a higher angular resolution usually implying a poorer sensitivity. Given the beam sizes and rms noises obtained from different bands at different weightings, we concluded that the best compromise between resolution and sensitivity is obtained by convolving all images to a circular beam with a size of 50~m.a.s., equivalent to a physical size of 7.35~au at the distance of 147 pc (Galli et al. 2018). Then, we obtained cleaned images at each wavelength with angular resolutions below 50 m.a.s. and convolved them by using the task \emph{imsmooth} from CASA.

 The final ALMA images at bands 7 and 6 used in our analysis are simply the reference images downloaded from the ALMA archive and convolved to a circular beam of 50 m.a.s. For ALMA band 4 and VLA data, we used images made with values of the parameter robust that gave us the highest sensitivity with an angular resolution just below 50~m.a.s. and then convolved these images. Since the S/N ratio in the VLA bands is low after some tens of au from the center of the disk, we obtained a higher sensitivity image by combining the free-free subtracted Ka and Q band data in a single image which corresponds to a central frequency of 38~GHz and a wavelength of 8.0 mm. We do not use the K band data for the analysis of the dust emission since the jet emission is likely to be affecting a larger part of the disk center at this band (see Appendix \ref{ObsFreeFree} for details). Moreover, the dust emission at K band is weak, and the signal-to-noise is poor, and adding it to the analysis does not help, but results in larger uncertainties. All final images used in our analysis are shown in Figure \ref{Fig2} and parameters are given in Table \ref{Tab2}.
 
 We obtained radial profiles of the emission at the different wavelengths by averaging emission in elliptical rings with a width of 10~m.a.s. and whose excentricity is given by the parameters of inclination\footnote{The inclination angle is i=0$^\circ$ for a face-on disk and i=90$^\circ$ for an edge-on disk.} i=46.72$^\circ$ and P.A.=138.02$^\circ$ of the HL~Tau disk (ALMA Partnership et al 2015b). The intensity at each radius is given by the average intensity in the ring, and the errors are calculated as $\Delta \rm{I}_\nu = \rm{rms}_\nu / ( \Omega_{\rm ring} / \Omega_{\rm beam} )^{0.5}$, where $\Omega_{\rm ring}$ and $\Omega_{\rm beam}$ are the solid angles of the ring and the synthesized beam, respectively. We also converted the average intensities at each radius to brightness temperatures using the full Planck expression (not the Rayleigh-Jeans approximation) to avoid errors at short wavelengths at the external parts of the disk where the dust temperatures are expected to be low. From the intensities at different wavelengths, we obtained radial profiles of spectral indices combining different wavelengths: $\alpha$(1.3 - 0.9 mm), $\alpha$(2.1 - 1.3 mm), and $\alpha$(8.0 - 2.1 mm). The radial profiles of brightness temperatures and spectral indices are shown in Figure \ref{Fig3}.

\section{Methods: Estimating Particle Sizes from Dust Opacity}

\begin{figure}[h]
\begin{center}
\includegraphics[width=\textwidth]{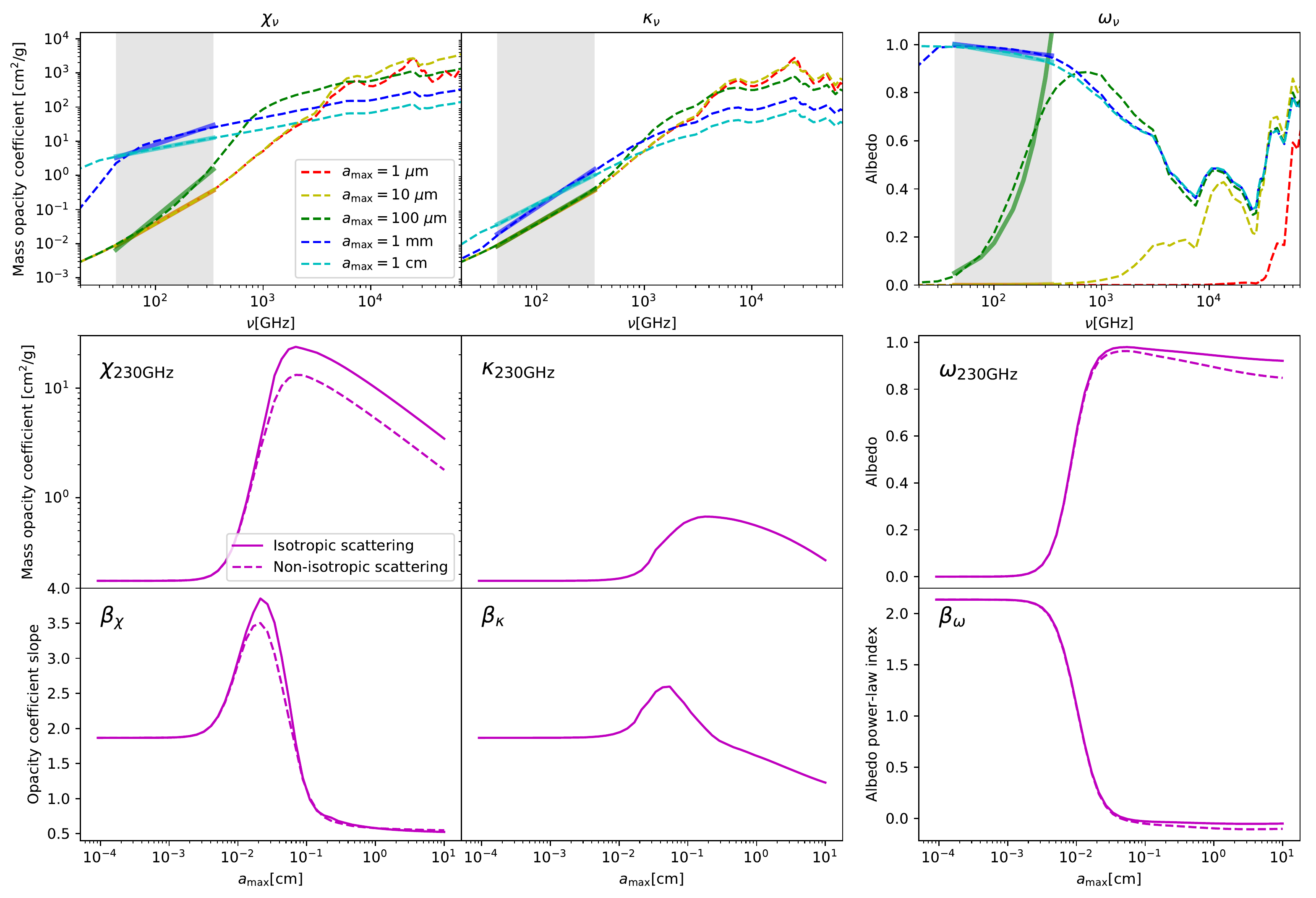}
\caption{\textbf{Top panels:} The spectral behaviour of the mass extinction coefficient ($\chi_\nu$), the mass absorption coefficient ($\kappa_\nu$) and the albedo ($\omega_\nu$), for particle sizes distribution with $p=3.5$ and different values of $a_{\rm max}$. Dust properties are computed using the D'Alessio et al. (2001) code for compact spherical grains and a dust mixture of silicates ($\rho_{\rm sil} = 3.3$ g cm$^{-3}$), organics ($\rho_{\rm org} = 1.5$ g cm$^{-3}$), and ice ($\rho_{\rm ice} = 0.92$ g cm$^{-3}$), with relative abundances of 26\%, 31\%, and 43\% respectively (Pollack et al. 1994). The vertical grey areas in all panels mark the wavelength range between 1 mm and 1 cm. Color solid lines are parametrizations of these coefficients as $\chi_\nu=\chi_0 (\nu/\nu_0)^{\beta_\chi}$, $\kappa_\nu=\kappa_0 (\nu/\nu_0)^{\beta_\kappa}$ and $\omega_\nu=\omega_0(\nu/\nu_0)^{\beta_\omega}$. \textbf{Central and Bottom panels:} Dependence of $\chi_\nu$, $\kappa_\nu$ and $\omega_\nu$ with $a_{\rm max}$. Solid lines are the values of the coefficients in the isotropic scattering approximation. Dashed lines are the values after correcting isotropic scattering effects through the assymmetry parameter $g$ assuming compact and spherical grains (see \S\ref{AbsSca} for details). The central panels show the dependence with $a_{\rm max}$ of the coefficients at a reference frequency of 230 GHz. Bottom panels show the values of the slopes describing the spectral variation of the coefficients with $a_{\rm max}$.}
\label{Fig4}
\end{center}
\end{figure}

 A method frequently used to estimate dust particle sizes in the interstellar medium and protoplanetary disks is through measurements of the dust opacity at millimeter wavelengths. Dust particles are usually assumed to follow a power-law size distribution in the form

\begin{equation} \label{partdist}
n(a) \propto  a^{-p} \quad ; \quad a_{\rm min} < a < a_{\rm max},
\label{partdist}
\end{equation}

\noindent where $n(a)da$ is the number of dust particles per unit volume with a size between $a$ and $a+da$, and $a_{\rm min}$ and $a_{\rm max}$ are the sizes of the smallest and the largest dust particles in the volume, respectively. The value of $a_{\rm min}$ has little effect on the opacity at millimeter wavelengths, and it is assumed to be 0.05 $\mu$m. In contrast, the value of $a_{\rm max}$ has a strong impact on the opacity of the emission at different wavelengths in the millimeter range (e.g., Miyake \ Nakagawa 1993).

 The spectral energy distribution (SED) of the dust emission is given by,

\begin{equation}
I_\nu = S_\nu [1 - \exp(-\tau_\nu )],
\end{equation}

\noindent where $I_\nu$, $S_\nu$ and $\tau_\nu$ are the emergent intensity, the source function and the optical depth at the observed frequency $\nu$, respectively. The optical depth depends on the dust surface density $\Sigma_{\rm dust}$ and the dust mass extinction coefficient $\chi_\nu$ which, in general depends on two effects: absorption and scattering. Thus, the extinction coefficient $\chi_\nu$ will be the sum of the absorption and the scattering coefficients, $\kappa_\nu$ and $\sigma_\nu$, respectively. The total optical depth can be written in several ways:

\begin{equation} \label{optdepth}
\tau_\nu = \Sigma_{\rm dust} \chi_\nu = \Sigma_{\rm dust} (\kappa_\nu + \sigma_\nu)  = \tau_\nu^{\rm abs} + \tau_\nu^{\rm sca}.
\end{equation}

 It is useful to define the albedo as the ratio between the opacity due to scattering and the total opacity,
 
\begin{equation}
\omega_\nu = \frac{\sigma_\nu}{\kappa_\nu + \sigma_\nu} = \frac{\tau_\nu^{\rm sca}}{\tau_\nu^{\rm abs}+\tau_\nu^{\rm sca}}.
\end{equation}

 In Figure \ref{Fig4} we show the spectral behaviour of the dust mass extinction coefficient, the absorption coefficient and the albedo for different particle sizes distributions. The values of the extinction and the absorption coefficients always increase with frequency at millimeter wavelengths, but the slope of the spectral variation of both coefficients is different for different values of $a_{\rm max}$ (see Figure \ref{Fig4}). The spectral behaviour of the albedo is more complex, but still, its average slope strongly depends on the maximum particle size. The optical depth at each wavelength will be determined by the spectral behaviour of these coefficients. Thus, by measuring the spectral behaviour of the optical depth, for example through observations at several wavelengths, it is possible to estimate the maximum particle size.
 
 In the following, we discuss how the properties of the dust opacity are usually used to easily estimate dust particle sizes by assuming that the dust opacity is dominated only by absorption. We discuss that this approximation, although valid for small particles like those found in the interstellar medium, it could not be correct in the case of protoplanetary disks. Then, we describe a more proper way to analyze millimeter data at several wavelengths to infer the maximum particle size by using a proper source function which also takes into account scattering effects. Finally, we apply it to the case of the HL~Tau disk.

\subsection{Interstellar Medium: Absorption-dominated dust opacity} \label{absonly}

 In cases where the dust opacity (equation \ref{optdepth}) is dominated only by absorption, the source function takes a very simple form which, in principle, allows to very easily estimate the maximum dust particle size directly from observations at only two wavelengths. In cases where the dust temperature is not expected to vary significantly along the line of sight (for example nearly face-on disks or those where the dust has effectively settled in the midplane), the source function can be approximated by the Planck function, $B_\nu (T_{\rm dust})$. At millimeter wavelengths, the spectral behaviour of the absorption coefficient is usually parametrized as $\kappa_\nu \propto \nu^{\beta_\kappa}$ (see Figure \ref{Fig4}). The slope of this coefficient, $\beta_\kappa$ is usually called in other works simply $\beta$. However, we use the nomenclature $\beta_\kappa$ to remark that this is related only to the spectral behaviour of the absorption coefficient and to differentiate later from the slope of the extinction coefficient. Thus, the optical depth will also follow a similar spectral behaviour than the absorption coefficient, i.e.,  $\tau\propto\nu^{\beta_\kappa}$. With this, the SED of the dust emission can be written as

\begin{equation} \label{SEDabs}
I_\nu =  B_\nu (T_{\rm dust}) [1 - \exp (-\tau_0 (\nu/\nu_0)^{\beta_\kappa})],
\end{equation}

\noindent where $\tau_0$ is the optical depth at a reference frequency $\nu_0$. In the Rayleigh-Jeans (R-J) approximation ($T_{\rm dust} >> 5 [\nu/100 GHz]$ K), which is valid for most cases in the millimeter wavelength range, equation \ref{SEDabs} simplifies for the optically thick ($\tau_\nu>>$1) and optically thin ($\tau_\nu<<$1) cases to:
 
\begin{eqnarray}
I^{\rm thick}_\nu \simeq I_0 (\nu/\nu_0)^{2} \quad ; \quad I^{\rm thin}_\nu \simeq I_0 (\nu/\nu_0)^{2+\beta_\kappa},
\end{eqnarray}
 
\noindent i.e., the spectral index of the emission between two millimeter wavelengths, $\alpha=log(I_{\nu1}/I_{\nu2})/log(\nu1/\nu2)$, will adopt values between $\alpha_{\rm thick}$=2 and $\alpha_{\rm thin}=2+\beta_\kappa$, depending on the optical depth of the emission at the observed wavelengths. Therefore, in the optically thin case, the value of $\beta_\kappa$ can be estimated from the spectral index at millimeter wavelengths as $\beta_\kappa\simeq\alpha-2$. The maximum particle size can be inferred directly from the measured value of $\beta_\kappa$ (see Figure \ref{Fig4}). This is the most usual method to estimate dust particle sizes in the interstellar medium and protoplanetary disks. However, the initial assumptions, optically thin emission and dust opacity dominated by absorption at millimeter wavelengths, might not be appropriate in the case of protoplanetary disks. 

 First, one can assume that $\beta_\kappa\simeq\alpha-2$ only in the case in which the spectral index has been obtained from optically thin wavelengths, i.e., $\tau_\nu<<$1 for both wavelengths involved in the calculation of the spectral index $\alpha$. In any other case, even for moderately optically thin emission with $\tau_\nu<$1, but near 1, we will obtain that $\alpha-2<\beta_\kappa$, which is most probably the actual case in several regions of the protoplanetary disks. As commented above, the most recent high angular resolution observations of protoplanetary disks with ALMA have revealed that emission at the central parts of the disks and some of the bright rings, could be very optically thick even at long wavelengths (e.g., Pinte et al. 2016, Jin et al. 2016, Liu et al. 2017). This makes extremely difficult to study the dust properties in these regions even with ALMA and could easily lead to overestimations of the particle sizes. For example, in a disk containing only small particles with sizes $\lesssim$~100~$\mu m$, the slope of the absorption coefficient is $\beta_\kappa\simeq2$ (see Figure \ref{Fig4}). In the absorption-only approximation, we should expect spectral indices $\alpha\simeq4$ in the optically thin regions, such as the low density gaps or the external parts of the disk. However, optically thick regions will show spectral indices $\alpha\simeq$2, which could wrongly be interpreted as $\beta_\kappa\simeq0$ and the presence of centimeter-sized particles (e.g., Galv\'an-Madrid et al. 2018; see also Figure \ref{Fig4}).
 
 Second, ignoring scattering effects could be even worse than the assumption of optically thin emission. Note that scattering can only be ignored in the case of very small particles ($\omega_\nu\simeq0$ only for $a_{\rm max}\lesssim$50 $\mu$m; see Figure \ref{Fig4}). However, in protoplanetary disks, we should expect millimeter- or even centimeter-sized particles, at least in some parts of the disk. But, millimeter-sized particles have large values of albedo at millimeter wavelengths ($\omega_{\nu} > 0.9$ for $a_{\rm max}\gtrsim$500 $\mu$m; see Figure \ref{Fig4}), and then, dust opacity will be indeed dominated by scattering. In this situation, the initial assumption is not fulfilled, and the main observational consequence is that the spectral index between two wavelengths will be not anymore easy to relate to the slope of the absorption coefficient (e.g., Sierra et al. 2019, Zhu et al. 2019). Instead, in the optically thin regime, the shape of the SED will be related to the spectral behaviour of the extinction coefficient $\chi_\nu$=$\kappa_\nu$+$\sigma_\nu$. Therefore, the parameter that shapes the SED is not the slope of the absorption coefficient, $\beta_\kappa$, but that of the extinction coefficient, $\beta_\chi$. Note that for $a_{\rm max}\gtrsim$500~$\mu$m, $\beta_\chi$ has lower values than $\beta_\kappa$ (see Figure \ref{Fig4}). Thus, again, ignoring scattering effects, could easily lead to overestimation of the dust particle sizes in protoplanetary disks. Moreover, ignoring scattering will also lead to underestimation of the dust mass because one will consider that opacity is dominated by the absorption coefficient which has lower values than the extinction coefficient (Zhu et al. 2019; see Figure \ref{Fig4}).
 
\subsection{Protoplanetary Disks: Scattering-dominated dust opacity} \label{AbsSca}

 From the discussion above, it is clear that a proper analysis of the millimeter data in protoplanetary disks needs to take into account both, absorption and scattering effects in the dust opacity. Then, the source function in the radiative transfer equation should be rewritten to include a term due to scattering. At each radius in a protoplanetary disk, we can write:

\begin{equation}
S_\nu (T) = \omega_\nu J_\nu + (1-\omega_\nu) B_\nu(T),
\end{equation}

\noindent where $J_\nu$ is the local mean intensity. In the case of a vertically isothermal slab and isotropic scattering, $J_\nu$ can be approximated by the analytical solution found by Miyake \& Nakagawa (1993) for a vertically isothermal slab:

\begin{equation}
J_\nu=B_\nu (T) [ 1+ f(t,\tau_\nu,\omega_\nu) ],
\end{equation}

\noindent where

\begin{equation}
f(t,\tau_\nu,\omega_\nu)=\frac{\exp(-\sqrt{3}\epsilon_\nu t) + \exp(\sqrt{3} \epsilon_\nu(t-\tau_\nu))}{\exp(-\sqrt{3}\epsilon_\nu\tau_\nu)(\epsilon_\nu-1)-(\epsilon_\nu+1)}.
\end{equation}

 The optical depth is now given by $\tau_{\nu}=\Sigma_{\rm dust} \chi_\nu$, and $t$ is a variable optical depth, both measured perpendicular to the disk mid plane, and 

\begin{equation}
\epsilon_\nu = \sqrt{1-\omega_\nu}.
\end{equation}

 In order to account for inclination effects, we have to also correct the optical depth by the inclination angle ($i$) of the disk by integration along $\tau_\nu/\mu$ , where $\mu=cos (i)$. With all this, the emergent specific intensity is obtained by (Sierra et al. 2019):

\begin{equation} \label{SEDsca}
I_\nu=\int_{0}^{\tau_\nu/\mu} S_\nu(T) e^{-t/\mu} \frac{dt}{\mu} = B_\nu(T) [(1-\exp(-\tau_\nu/\mu)) + \omega_\nu F(\tau_\nu,\omega_\nu)],
\end{equation}

\noindent where

\begin{eqnarray}
F(\tau_\nu,\omega_\nu)= \frac{1}{\exp(-\sqrt{3}\epsilon_\nu\tau_\nu)(\epsilon_\nu-1)-(\epsilon_\nu+1)} \times \nonumber \\
\times \left[ \frac{1-\exp(-(\sqrt{3}\epsilon_\nu+1/\mu)\tau_\nu)}{\sqrt{3}\epsilon_\nu\mu +1}
+\frac{\exp(-\tau_\nu/\mu)-\exp(-\sqrt{3}\epsilon_\nu\tau_\nu)}{\sqrt{3}\epsilon_\nu\mu -1} \right].
\end{eqnarray}

A similar equation derived using Eddington-Barbier approximation is given in Zhu et al. (2019). Now, the optical depth depends on the extinction coefficient $\chi_\nu$. In a similar way than it is usually done for the absorption coefficient, the spectral behaviour of the extinction coefficient $\chi_\nu$ and the albedo $\omega_\nu$ can be parameterized in the millimeter wavelength range as,
 
\begin{equation}
\chi_\nu \propto \nu^{\beta_\chi} \quad ; \quad \omega_\nu \propto \nu^{\beta_\omega}
\end{equation} 

 Note that now, the changes in optical depth with frequency will be given by the slope of the extinction coefficient, i.e.,
 
\begin{equation}
\tau_\nu = \tau_0 (\nu/\nu_0)^{\beta_\chi}.
\end{equation}

Note that these equations assume isotropic scattering, which may be a bad approximation for $2\pi a \gg \lambda$. The impact of this approximation can be reduced by replacing in all equations the scattering coefficient $\sigma_\nu$ by  an effective scattering coefficient in the form
 
\begin{equation}
\sigma_\nu^{eff} = (1-g_\nu) \sigma_\nu,
\end{equation} 

\noindent where $g_\nu$ is the assymmetry parameter, i.e., the expectation value of cos $\theta$, where $\theta$ is the scattering angle (e.g. Ishimaru 1978, Birnstiel 2018). We used the dielectric optical constants to compute the value of $g$ for compact and spherical grains, following Kr\"ugel (2003). The scattering coefficient $\sigma$ is first corrected for each grain size, and then  they are averaged for a given particle size distribution. In Figure \ref{Fig4}, we show that this correction slightly modify the values and spectral behaviours of the extinction coefficient and the albedo. 

As can be seen, equations including scattering effects in the dust opacity are much more difficult to use for the interpretation of the observational data. In particular, the spectral index of the emission between only two wavelengths is not easy to relate anymore with the dust opacity and the dust properties (e.g., Liu et al. 2019, Zhu et al. 2019). Equation \ref{SEDsca} depends on 6 parameters: $T_{\rm dust}$, $\Sigma_{\rm dust}$, $\chi_0$, $\beta_\chi$, $\omega_0$, and $\beta_\omega$. However, note that given a particle size distribution ($a_{\rm max}$ and $p$) and grain composition, the values of $\chi_0$, $\beta_\chi$, $\omega_0$ and $\beta_\omega$ are fixed. The parameters defining the particle size distribution, $p$ and $a_{\rm max}$ are strongly degenerated. However, we still can assume a value for one of them and leave the other as a free parameter. The most convenient is to assume a value for $p$, and find the value of the maximum particle size. In this way, equation \ref{SEDsca} only depends on 3 free parameters: $T_{\rm dust}$, $\Sigma_{\rm dust}$, $a_{\rm max}$. Therefore, we should be able to properly estimate the dust properties in a protoplanetary disk, including scattering, regarding that observations at 4 different wavelengths are available, which is the case of HL~Tau.
 
\subsection{Radial SED Fitting} \label{SEDfit}

 The images described in \S\ref{FinalImages} cover a wide range of wavelengths, from ~0.9~mm to ~8.0~mm, which implies that at each radius we detect emission with very different optical depths. Therefore, we can perform a radial analysis of the SED of the dust emission using equation \ref{SEDsca} to simultaneously obtain the dust temperature, the dust surface density and the maximum particle size at each radius. Since we are not assuming any value of the optical depth at any wavelength, and we are including scattering in the dust opacity, we are avoiding effects that usually result in overestimation of the particle size (see discussion on \S\ref{absonly}). Note that the main assumption in equation \ref{SEDsca} is a constant dust temperature in the line of sight. This assumption is a reasonable approximation in the case of disks where the dust is already well settled towards the mid-plane, which is the case of HL Tau (the scale height in HL Tau is estimated to be only $\sim$1~au at a radius of $\sim$100 au; Pinte et al. 2015).

 To explore the dust properties radially in HL Tau, we assume a value of $p=3.5$ of the coefficient of the particle distribution (equation \ref{partdist}), and we consider a reference frequency $\nu_0$=230 GHz, which corresponds to a wavelength of 1.3~mm. The dust opacity properties are computed using the D'Alessio et al. (2001) code for compact spherical grains and a dust mixture of silicates ($\rho_{\rm sil} = 3.3$ g cm$^{-3}$), organics ($\rho_{\rm org} = 1.5$ g cm$^{-3}$), and ice ($\rho_{\rm ice} = 0.92$ g cm$^{-3}$), with relative abundances of 26\%, 31\%, and 43\% respectively (Pollack et al. 1994) (see Figure \ref{Fig4}). For the fit of the SED at each radius, we use the function \emph{curvefit}\footnote{https://docs.scipy.org/doc/scipy/reference/generated/scipy.optimize.curve\_fit.html} in \emph{python}, which use non-linear least squares methods to fit a function to a set of data. 
 
 At each radius, we follow an iterative process. We first assume a particle size distribution with a initial value of the maximum particle size, $a^{\rm init}_{\rm max}$ which corresponds to initial values of $\omega^{\rm init}_0$ and $\beta^{\rm init}_\omega$. Then, we fit the SED to equation \ref{SEDsca} to obtain values of $T_{\rm dust}$, $\tau_0$, and $\beta_{\chi}$. The obtained value of $\beta_{\chi}$ corresponds to a new value of $a_{\rm max}$, and then to new values of $\omega_0$ and $\beta_\omega$. If these are different to those assumed initially, we fit again using the new values of $\omega_0$ and $\beta_\omega$ to obtain a new set of $T_{\rm dust}$, $\tau_0$, $\beta_{\chi}$ and $a_{\rm max}$. We repeat this until $a_{\rm max}$ converges (i.e., differences are lower than 5\%) which typically occurs in 2 or 3 iterations and the result is independent of the initial value of $a_{\rm max}$ used. In order to estimate errors, at each radius, we repeat this procedure 1000 times, each time we randomly vary intensities at the different wavelengths within its error bars following a normal distribution with $\sigma$ equal to the error. Thus, at each radius, the nominal value of a parameter is defined as the mean and the error as the standard deviation of the 1000 fits. 
 
\begin{figure}[t]
\begin{center}
\includegraphics[width=\textwidth]{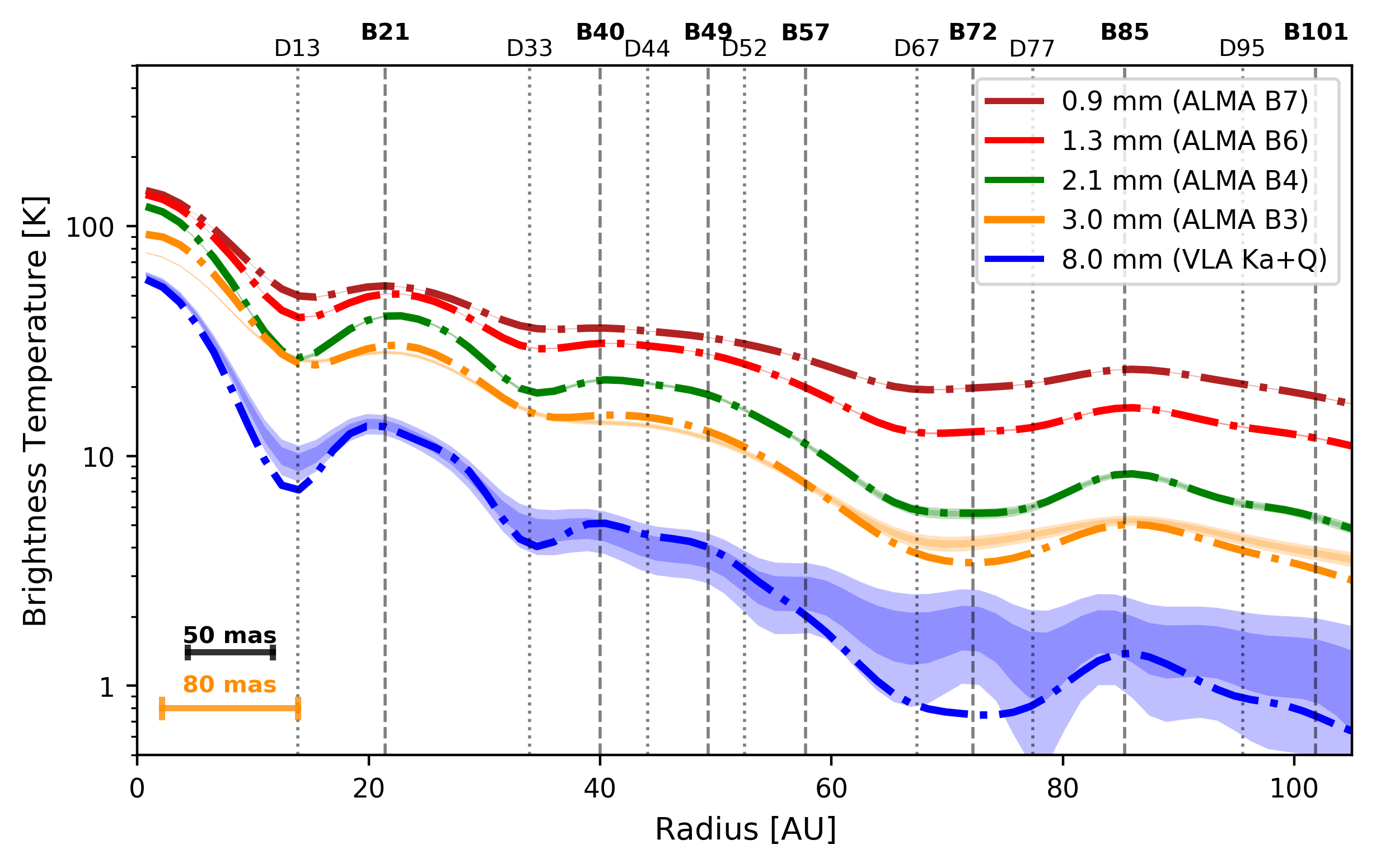}
\caption{Comparison between the predicted brightness temperatures at each wavelength from our absorption+scattering model, with the VLA and ALMA observations. Predictions from the model are shown as dashed-pointed lines. For the observations we show uncertainties at 1- and 2-$\sigma$ levels. For our analysis, we only take into account data with the highest angular resolution (50 m.a.s.; ALMA bands 7, 6, and 4 and VLA Ka+Q band). We also include a comparison of the prediction of our model with a lower angular resolution ($\sim$80 m.a.s.) ALMA data at 3.0 mm (band 3) after convolution of our results at this wavelength.}
\label{Fig5}
\end{center}
\end{figure}

 In Figure \ref{Fig5}, we show a comparison between the observed brightness temperatures at each wavelength, and the predictions from our best fit model. We found a good agreement between model and observations at all wavelengths. There is another available ALMA image of HL Tau at 3.0 mm, also part of the Science Verification program, which was previously presented by ALMA Partnership et al. (2015b), but with lower angular resolution ($\sim$80 m.a.s.) than the data used in our analysis. We convolved the prediction of our model at 3.0 mm to the angular resolution of this image, and compare the radial profiles of the brightness temperature. As can be seen in Figure \ref{Fig5}, there is also good agreement between these data and our model.

 The procedure described above allows us to obtain at each radius $T_{\rm dust}$, $\tau_0$, $\beta_\chi$, $\omega_0$, $\beta_\omega$, and $a_{\rm max}$. The last four parameters, which are not independent, are consistent between them after our iterative procedure, and they correspond to a single value of the extinction coefficient, $\chi_0$. Thus, we can also obtain the dust surface density as $\Sigma_{\rm dust}=\tau_0/\chi_0$. In this way, the final result from our analysis is to obtain radial profiles of $T_{\rm dust}$, $\Sigma_{\rm dust}$, and $a_{\rm max}$. 
 
\section{Results and Discussion}

\begin{figure}[p]
\begin{center}
\includegraphics[width=0.9\textwidth]{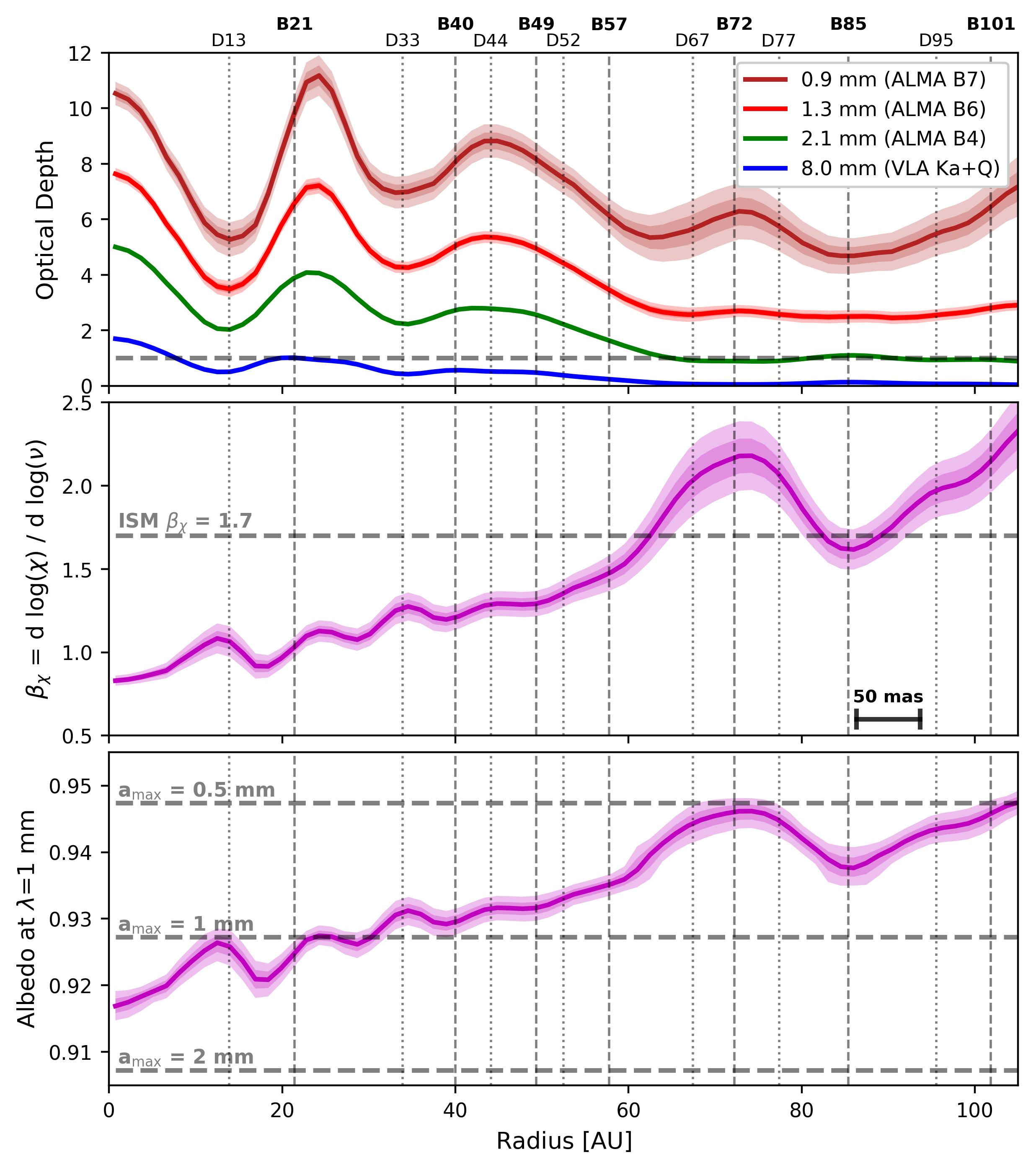}
\caption{\footnotesize{Results of the radial SED fitting (see \S\ref{SEDfit} for details). Vertical lines in all panels show the positions of bright and dark rings with a notation following Huang et al. (2018). \textbf{Top panel:} Total (scattering+absorption) optical depth profiles at the different observed wavelengths. The horizontal dashed line marks the limit between optically thick and optically thin emission, $\tau$=1. \textbf{Central panel:} Radial profile of the slope of the extinction (scattering+absorption) coefficient, $\chi_\nu$ = $\kappa_\nu$ + $\sigma_\nu$ $\propto$ $\nu^{\beta_\chi}$. This parameter describes the spectral behaviour of the extinction coefficient in the 1-10 mm wavelength range. An horizontal dashed line shows the value observed in the interstellar medium ($\beta_\chi$=1.7). \textbf{Bottom panel:} Radial profile of the albedo at a wavelength of 1~mm. Horizontal dashed lines show the expected values for different values of the maximum size in the dust particle distribution. In all panels, errors are shown at the 1- and 2-$\sigma$ levels as filled bands around the nominal values.}}
\label{Fig6}
\end{center}
\end{figure}

 In Figure \ref{Fig6}, we show the results for the parameters that describe the dust opacity: the optical depth at different wavelengths ($\tau_\nu$), the slope of the extinction coefficient ($\beta_\chi$), and the albedo at the reference wavelength ($\omega_0$). Note that the optical depth includes both effects, absorption and scattering, and then, the spectral behaviour of the optical depth is given by the slope of the extinction coefficient, i.e., $\beta_\chi$=$d \log (\chi_\nu)$/$d \log (\nu)$. As can be seen, at this angular resolution, emission at the ALMA wavelengths is optically thick $\tau\simeq1-10$ at all radii. Only the band 4 image is marginally optically thick ($\tau_\nu\sim 1$) at radii $>$60~au. The VLA image at 8.0~mm is marginally optically thick ($\tau\simeq1-2$) only at the center of the disk and at the position of the first bright ring B21, and it is optically thin for the rest. These results are consistent with the previous results from radiative transfer modeling (Liu et al. 2017). The radial gradients of $\beta_\chi$ and $\omega_0$ imply variations in $a_{\rm max}$ between $\sim$1.5~mm at the center of the disk and $\sim$0.5~mm at outer radii. 

\begin{figure}[p]
\begin{center}
\includegraphics[width=\textwidth]{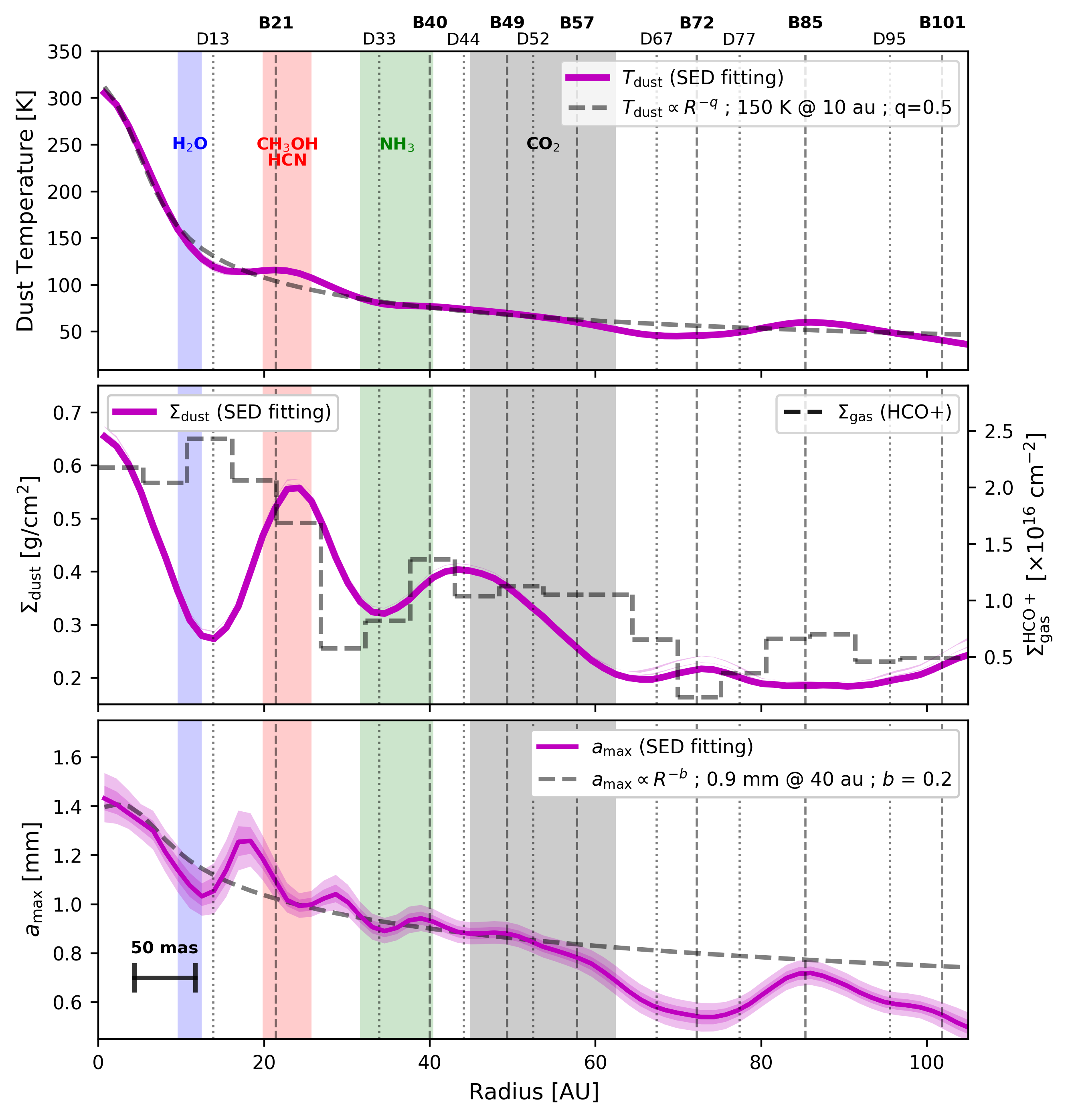}
\caption{The dust distribution in HL Tau derived from our radial SED fitting of the ALMA and VLA data. Vertical lines in all panels show the positions of bright and dark rings with a notation following Huang et al. (2018). \textbf{Top panel:} Dust Temperature radial profile. The dashed line is a power-law fit to the radial variation of the dust temperature. \textbf{Central panel:} Dust surface density. The dashed grey line is the gas surface (number) density obtained by Yen et al. (2016) from HCO$^+$ observations. \textbf{Bottom panel:} Radial profile of the maximum dust particle size obtained from our SED fitting. The grey dashed line is power-law fit to the radial variation of $a_{\rm max}$. In all panels, vertical color bands mark the positions of snow lines of several molecules according to the dust temperature power-law profile obtained from our fit (ranges of freezing temperatures for these molecules are taken from Zhang et al. 2015). Errors are shown at the 1- and 2-$\sigma$ levels as filled bands around the nominal values.}
\label{Fig7}
\end{center}
\end{figure}

 In Figure \ref{Fig7}, we show the results for the parameters describing the physical conditions of the dust in the disk: dust temperature ($T_{\rm dust}$), dust surface density ($\Sigma_{\rm dust}$) and the maximum particle size ($a_{\rm max}$). The values of $T_{\rm dust}$ were obtained from a fit to the emission convolved to the resolution of the images, i.e., at each radii we are actually obtaining an effective temperature of a region with a size of $~$7.35~au. Then, in Figure \ref{Fig7} we compare with a power-law temperature distribution convolved to the same resolution than our observations. As can be seen, the dust temperature in the HL~Tau disk approximately follows a power-law in the form $T_{\rm dust}=150 \times [R/10 AU]^{-0.5}$ K. Note that, from this dust temperature profile, we can predict the locations of the snow lines of different molecules, which are shown in Figure \ref{Fig7} as vertical color bands (ranges of freezing temperatures for these molecules are taken from Zhang et al. 2015). The dust surface density and the maximum particle size also decrease with radius (see Figure \ref{Fig7}). The total dust mass of the disk in this model is $\sim$1.0$\times$10$^{-3}$~M$_\odot$ which is consistent with previous modeling of the HL Tau disk (e.g., D'Alessio et al. 2001; Pinte et al. 2016; Liu et al. 2017). 

 In Appendix \ref{comparison} we show a comparison of our results with those that would be obtained assuming that opacity is dominated only by absorption. As commented in previous subsections, the usual approximation of opacity dominated by absorption, leads to important errors in the inferred dust properties. We then emphasize the necessity of including also the effects of scattering when analyzing millimeter observations of protoplanetary disks.
 
 The high angular resolution of our analysis allows us to detect local variations associated with the presence of dark and bright rings in the disk. In the following, we discuss both, the global behaviour and the local variations of the dust properties. 

\subsection{The Global Particle Dust Distribution in HL Tau}

 Our results imply that dust particles in the HL~Tau disk have grown up to, at least, millimeter sizes. Maximum particle sizes are $\sim$1.5~mm at the center of the disk and decrease to $\sim$0.5~mm at the external disk (see Figure \ref{Fig7}). This is not consistent with the recent results obtained through dust polarization measurements which were interpreted as a disk with a population of only submillimeter particles, and a maximum grain size of $\sim$100~$\mu$m for the whole disk (Kataoka et al. 2017). In contrast, we find that only the most external parts of the disk (R$>$60~au) are compatible with particle distributions with $a_{\rm max}<$1~mm, but still, these are 5 times larger than estimated from polarization results. An exploration of possible errors as a consequence of flux calibration uncertainties, neither decrease the values of $a_{\rm max}$ (see Appendix \ref{FluxUnc}). In our analysis, dust opacity includes scattering and absorption and we are not assuming optically thin emission at any wavelength. Therefore, we are avoiding overestimating the maximum particle size due to the usual assumptions. Note that the value of the extinction coefficient for $a_{\rm max}$=100~$\mu$m is $\sim$50 times smaller than the value for millimeter particles (see Figure \ref{Fig4}). Therefore, a disk made with such small particles should be $\sim$50 times more massive in order to explain the optical depths found in our analysis. This would imply that the total mass of the disk (gas+dust) would be around $\sim$1~M$_\Sun$, similar to that of the central star and would imply that the disk is extremely unstable. To explain a disk made of small particles with a reasonable total dust mass, it would be then necessary to decrease the optical depth at all wavelengths by a factor of $\sim$50, which will make the emission optically thin at all wavelengths (see Figure \ref{Fig6}). But then, the dust temperature should be increased by a similar factor, sublimating all the dust in the disk. Therefore, we find very difficult to explain all the available millimeter observations in HL Tau with a disk made of particles with sizes $\lesssim$100 $\mu$m.
 
 One possibility to explain the discrepancy in particle sizes between both methods could be the likely presence of several dust grain populations with different $a_{\rm max}$ (Kataoka et al. 2017, Liu et al. 2017); the relative location of each particle population in the disk, together with the wavelengths and angular resolutions used in the polarization observations, could produce a polarization pattern similar to that obtained in a disk with only submillimeter particles. For example, while the largest particles should be located only in the mid-plane of the disk, the disk could still have an atmosphere with very small particles which could dominate the dust self-scattering polarization observed at short wavelengths and low angular resolution with ALMA.

We compare our results with those obtained in disks around similar stars. There is a handful of disks around T-Tauri stars where particle size distributions have been studied through high angular resolution observation in a similar wavelength range than our study: AS209, CQ Tau, CY~Tau, DoAr25, FT~Tau, and DR~Tau (P\'erez et al. 2012, 2015; Trotta et al. 2013; Tazzari et al. 2016). In order to compare with our results, we fit the obtained particle size distribution to a power-law variation of the maximum particle size with radius in the form $a_{\rm max} = a_0 [R/R_0]^{-b}$ (similar to the analysis performed for other disks around T-Tauri stars; e.g., Tazzari et al. 2016). The global gradient in $a_{\rm max}$ in HL~Tau approximately follows a power-law with $a_0\sim$0.9~mm at 40~au, and $b$=0.2 (see Figure \ref{Fig7}). In CY~Tau and DoAr 25 radial distributions consist of $a_{\rm max}\simeq$1~mm for most of the disk, slowly increasing to internal radii, and then a very steep increase to values of several centimeters for radii below 20~au (P\'erez et al. 2015). In the case of CQ~Tau, AS209, FT~Tau, and DR~Tau, the radial variations of $a_{\rm max}$ obtained for the case of $p$=3.5, were parameterized as power-laws with values of $a_0\simeq$4-8~mm (at 40~au), and $b\simeq$1-2 (Trotta et al. 2013; Tazzari et al. 2016). All these previous results differ from our results in HL~Tau: we found smaller particles in the whole disk ($a_0\simeq$0.9~mm at 40~au), the external parts of the HL~Tau disk are indeed compatible with only submillimeter particles ($a_{\rm max}\simeq$500~$\mu$m), the global gradient in $a_{\rm max}$ is flatter ($b\simeq$0.20), and we do find neither a steep increase in $a_{\rm max}$, nor centimeter-sized particles at the center of the disk. If we take into account potential flux calibration errors, we found that the millimeter data in HL~Tau could be consistent with $a_{\rm max}\simeq$2.0~mm at most, at the center of the disk (see Appendix \ref{FluxUnc}), which is still very far from centimer-sized particles.
 
 The large values of $a_{\rm max}$ obtained in other T-Tauri disks is most probably a consequence that previous studies usually ignore scattering effects in their analysis. As we have discussed, this assumption could easily lead to overestimations of the dust particle sizes in protoplanetary disks (see \S\ref{absonly}). This is especially important in the central parts of the disks where optically thicker emission is expected. Indeed, in the case of HL Tau, we also note that repeating our analysis ignoring the scattering effects lead to very large sizes for the particles in the inner half part of the disk ($a_{\rm max}>>$1~cm for R$<$50~au; see Appendix \ref{comparison}).

 Our results in HL~Tau put an important constraint for the dust grain growth process: millimeter-sized particles are created before the first million years in the evolution of T-Tauri disks, at least in those disks which already show radial substructures. We emphasize the necessity to extend this kind of studies, which model the SED in a wide range of millimeter wavelengths and which includes the effects of scattering, to other disks with different ages. Also, in order to understand the role of rings and gaps in the dust evolution, it is important to study disks not showing radial substructures. 

\subsection{On the Nature of Bright and Dark Rings} \label{nature}
 
 The nature of bright and dark rings is being intensively debated in the last years. On one hand, it has been long expected to detect these kind of structures as a consequence of already formed (proto)planets in the disk (e.g., Zhu et al. 2014; P\'erez et al. 2015). On the other hand, we know that these structures can also be formed by other mechanisms not requiring the presence of protoplanets, such as magnetized disks (Flock et al. 2015), fast pebble growth near condensation fronts (Zhang et al. 2015), and sintering-induced dust rings (Okuzumi et al. 2016, 2019; Hu et al. 2019). Irrespective of their origin, once formed, dense rings are excellent places to accumulate and concentrate large dust grains (e.g., Zhu et al. 2014; Flock et al. 2015; Ruge et al. 2016; Sierra et al. 2017, 2019), and their ubiquitous presence in disks of all ages (e.g. van der Marel et al. 2019) have lead to propose that the formation of rings is a necessary first step for the formation of terrestrial planets (e.g., Carrasco-Gonz\'alez et al. 2016). It is then fundamental to study the dust properties in the dark and bright rings in order to discern how they were formed and how they will evolve.

 The high angular resolution and the wide wavelength range of our analysis of the HL~Tau disk allows to study differences in the particle size distribution between bright and dark rings. We clearly see differences in the dust properties between dark and bright rings (see Figure \ref{Fig7}). Note that, in some cases, our angular resolution is very similar to the separation between radial features. The effect of this is to smooth differences in the obtained paramaters. Therefore, the contrast between density or particle sizes between dark and bright rings could actually be higher. We use all the available observational information in order to discuss the true nature of the features and to try to discern their formation mechanism. The positions of the previously identified dark and bright rings in HL~Tau are marked in Figure \ref{Fig7} as vertical dotted and dashed lines, respectively. We labeled them following the notation introduced by Huang et al. (2018) which consists of the letter D (=dark) or B (=bright) followed by the radius in au (see Figure \ref{Fig7}). As commented above, the dust temperature profile obtained from our modeling fixes the positions of the condensation fronts of several of the most important molecules present in protoplanetary disks. In Figure \ref{Fig7}, we show the positions of the snow lines of H$_2$O, CH$_3$OH, HCN, NH$_3$, and CO$_2$ as vertical color bands (ranges of freezing temperatures for these molecules are taken from Zhang et al. 2015). We also show in Figure \ref{Fig7} the gas (number) surface density obtained from HCO$^+$ observations (Yen et al. 2016).

 Many studies analyzing the presence of (proto)planets in HL~Tau have proposed a sub-Jupiter mass object located at an orbit of $\sim$70~au (e.g., Dipierro et al. 2015, Jin et al. 2016, Bae et al. 2017; Dong et al. 2018). This object would be the responsible for the opening of a wide gap between $\sim$60 and $\sim$85~au, which is the most prominent dark feature observed in the millimeter images (see Figures \ref{Fig1} and \ref{Fig2}), and encloses two dark rings (D67 and D77). It can be considered as a single, very wide, dark ring with a very narrow moderately bright ring (B72) at its center. The massive (proto)planet is usually proposed to be associated with the central bright ring. The D67-B72-D77 feature is also coincident with a gas gap detected in the HCO$^+$ observations (Yen et al. 2016). We find that the whole region between $\sim$60 and $\sim$85~au is also associated with very low dust densities, the bright ring B72 is moderately denser than the surroundings, and the whole region is also associated with smaller particle sizes than expected from the global gradient in $a_{\rm max}$ (see Figure \ref{Fig7}). All these results imply an important discontinuity in the disk around an orbit of $\sim$70~au, which would be consistent with the presence of a massive (proto)planet.
 
 While some studies invoke a single object at $\sim$70~au to explain the morphology of HL~Tau (e.g., Bae et al. 2017 and Dong et al. 2018), others also propose additional objects located at dark rings D13 and D33 (e.g., DiPierro et al. 2015 and Jin et al. 2016). The presence of a protoplanet at an orbit of $\sim$33~au is also supported by the gas gap seen in the HCO+, but a similar gas gap is not observed in D13 (Yen et al. 2016; see Figure \ref{Fig7}). We found that both dark rings, D13 and D33, are associated with low dust density and smaller particles (see Figure \ref{Fig7}). These results, in principle would support the idea of (proto)planets embedded in these positions. However, we also note that the expected positions of the snow lines of water and ammonia coincide with the positions of D13 and D33 (see Figure \ref{Fig7}), which suggest a different formation mechanism for D13 and D33 (see below). 
 
 It is known that freezeout or sublimation of gas volatiles on the surface of dust grains can significantly change the fragmentation and sticking properties of the particles (G{\"u}ttler et al. 2010). As a consequence, several models of dust evolution have predicted accumulations of large (e.g., Ros \& Johansen 2013, Pinilla et al. 2017) or small particles (e.g., Okuzumi et al. 2016, 2019) near the snow lines which could explain the radial features seen at millimeter wavelengths. We note that several snow lines coincide with the position of dark and bright rings (see Figure \ref{Fig7}). The dark rings D13 and D33 are associated with low dust density and small particles and they coincide with the H$_2$O and NH$_3$ snow lines, respectively. In contrast, the bright ring B21 is associated with high dust density, small particles and coincides with the position of the CH$_3$OH and HCN snow lines (see Figure \ref{Fig7}). The location of the CO$_2$ snow line could also have a role in the formation of the several bright and dark rings located between 40 and 60 au (see Figure \ref{Fig7}). These results strongly support the idea that the most internal radial features are not created by planet-disk interactions, but they are actually a consequence of changes in the dust properties related to the presence of the snow lines.

\newpage
\section{Summary and Conclusions}

 In this paper we presented an analysis of high quality ALMA and VLA images of HL~Tau covering a wide range of wavelengths, from 0.8~mm to 1~cm. The high angular resolution and sensitivity of these images allow us to investigate the properties of the dust with an unprecedented high resolution ($\sim$7.35~au). Other works studying the dust properties in T-Tauri disks usually do not take into account scattering effects and assume that emission is optically thin at millimeter wavelengths. We have discussed that, while these assumptions are valid for the interstellar medium, they might lead to incorrect results in the case of protoplanetary disks. In particular, they could easily lead to overestimations of the particle sizes and underestimation of the dust mass. We take advantage of the large number of high quality data available in the case of the HL~Tau disk to fit the millimeter SED at each radius to a more general equation including scattering. In this way, we are able to obtain radial profiles of the dust temperature, the dust surface density and the maximum particle size. We discuss our results in the context of grain growth and planet formation. Our main conclusions can be summarized as:
 
\begin{enumerate}

\item{} Dust particles have grown up to millimeters in size in the HL~Tau disk. We detect a global radial gradient in the maximum particle size. At the center, particles have maximum sizes around a few millimeters. The maximum particle size decreases with radius and it is slightly below 1~mm at 100~au.

\item{} We find differences in the dust properties between bright and dark rings. Dark rings are always associated with low dust density and small particles. Bright rings are always associated with high dust density and, except in one case, with large dust particles. The case of the most internal bright ring, B21 slightly differs: it is associated with high dust density and smaller particles.
 
\item{} Our results suggest different origins for different radial features in the HL~Tau disk. The wide dark ring D67-D77 is associated with dust and gas gaps and with very small particles. This result support the origin of this feature as due to planet-disk interactions, as proposed by several previous studies. In contrast, the most internal features, D13, B21 and D33, are associated with slightly smaller particles than their surroundings and their positions are coincident with those of the H$_2$O, CH$_3$OH, HCN, and NH$_3$ snow lines. Several bright and dark rings located between 40 and 60 au could be also associated with the condensation front of CO$_2$. This strongly supports that the radial features at radii $<$60~au in the HL~Tau disk are actually related to changes in the fragmentation and sticking properties due to the freezing or sublimation of molecules in the snow lines.

\item{} We discussed the errors in the inferred dust particles properties introduced by ignoring scattering effects. We emphasize that a correct analysis of millimeter data in protoplanetary disks should also include possible effects due to scattering. In this paper, we describe a simple procedure to more properly analyze millimeter images of protoplanetary disks. It is necessary to extend this kind of analysis to other disks with and without substructure and at different stages of evolution in order to understand grain growth and planet formation in protoplanetary disks.

\end{enumerate}

\noindent \emph{Acknowledgments.} This paper makes use of the following ALMA data: ADS/JAO.ALMA\#2011.0.0000X.SV15. ALMA is a partnership of ESO (representing its member states), NSF (USA) and NINS (Japan), together with NRC (Canada) and NSC and ASIAA (Taiwan), in cooperation with the Republic of Chile. The Joint ALMA Observatory is operated by ESO, AUI/NRAO and NAOJ. CC-G and RGM acknowledge support from UNAM-DGAPA PAPIIT Programmes IN108218 and IN104319. MF has received funding from the European Research Council (ERC) under the European Unions Horizon 2020 research and innovation programme (grant agreement No. 757957). ZZ acknowledges support from the National Science Foundation under CAREER Grant Number AST-1753168 and Sloan Research Fellowship. GA and MO acknowledge financial support from the State Agency for Research of the Spanish MCIU through the AYA2017-84390-C2-1-R grant (co-funded by FEDER) and through the ``Center of Excellence Severo Ochoa'' award for the Instituto de Astrof\'{\i}sica de Andaluc\'{\i}a (SEV-2017-0709). J.M.T. acknowledge financial support from the State Agency for Research of the Spanish MCIU through the AYA2017-84390-C2 grant (cofunded by FEDER). Software: CASA (McMullin et al. 2007). We thank an anonymous referee for a critical review of the manuscript and very useful suggestions.


\newpage 

\appendix

\section{Free-free substraction of VLA observations} \label{ObsFreeFree}

 It is well known that HL~Tau drives a powerful optical parsec scale jet (e.g, Anglada et al. 2007) and that significant contribution from free-free emission is expected at the VLA bands at the very center of the disk (Carrasco-Gonz\'alez et al. 2009, 2016). In our previous analysis of the Q band data, we used old archive data at lower frequencies to obtain a rough estimate of the free-free contribution by extrapolating the flux densities to 7.0 mm. From this, we concluded that free-free emission should be located $\lesssim$40~m.a.s. of the center of the disk, and that it accounts for 35-65\% of the total flux at 7.0~mm in this region (Carrasco-Gonz\'alez et al. 2016). Large uncertainties in this estimate are mainly due to very different angular resolutions between the low frequency data used for the extrapolation of the free-free emission and the 7.0~mm data (a factor $\sim$10 lower angular resolution at lower frequencies). Here, we take advantage of the availability of our new high quality data at several VLA bands to perform a more sophisticated subtraction of the free-free emission at the center of the disk.

\begin{figure}[h]
\begin{center}
\includegraphics[width=0.8\textwidth]{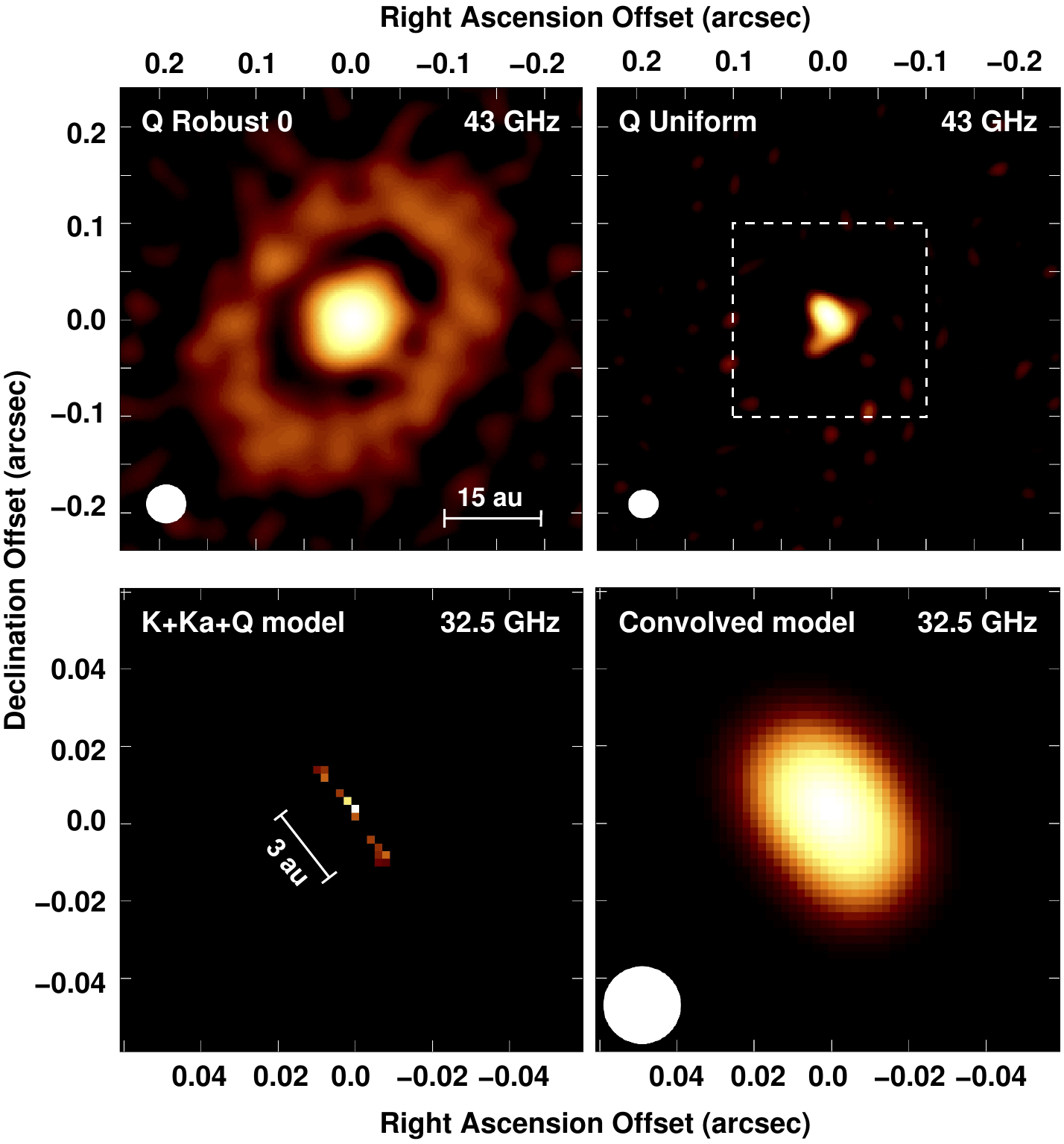}
\caption{Substraction of free-free emission at the center of the disk (see Appendix \ref{ObsFreeFree} for details). \textbf{Top Left:} VLA image at Q band with an angular resolution of $\sim$42 m.a.s. (robust=0 weighted). \textbf{Top Right:} VLA image at Q band with an angular resolution of $\sim$30 m.a.s. (uniform weighted). The bottom panels show the region enclosed in the dashed line square. \textbf{Bottom Left:} Deconvolved model components obtained from the MFS cleaning using the combination of K, Ka and Q bands data and uniform weighting. \textbf{Bottom Right:} Convolution of the deconvolved model in the left with a circular beam of 20 m.a.s.}
\label{FigA1}
\end{center}
\end{figure}

 We noted that the rms noise and the angular resolution obtained in the VLA data imply brightness limits to the detectable emission of 50, 70 and 100 K (for K, Ka and Q bands, respectively) for uniform weighted images. The expected brightness temperatures of the dust emission for all these bands is $\lesssim$60 K at the center of the disk (see Figure \ref{Fig3}). In contrast, we expect brightness temperatures of several hundreds of K due to the jet free-free emission, originated in material at temperatures of the order of several thousands of K. Therefore, the emission in uniform weighted maps should be dominated by free-free emission from the jet. This is indeed confirmed by the morphology seen in the Q band uniform weighted image (angular resolution $\sim$30~m.a.s.; see Figure \ref{FigA1}) which shows a very compact source elongated in the jet direction (NE-SW; P.A.=30$^\circ$) with a deconvolved size of $\sim$40~m.a.s. (from a Gaussian fit). Although this image might still contain some contribution from very bright dust emission at the very center of the source, its elongated morphology in the jet direction strongly suggests that it is dominated by the free-free emission from the jet. In the other bands, the angular resolution was not enough to resolve the jet morphology. However, we also expect that emission in uniform weighting maps at these bands is even more dominated by the jet emission since the brightness temperature of the dust at these wavelengths should be lower (intensity is weaker and optically thinner), and the brightness of the jet should be higher (the size of the optically thick region is larger) than at 7.0 mm. 
 
 To obtain a proper model of the jet emission, we made a single image combining all three VLA bands (K, Ka and Q) with the highest angular resolution possible ($\sim$30~m.a.s.; uniform weighting) and using multi-frequency synthesis (MFS) cleaning (Rau \& Cornwell 2011) which allow to simultaneousy model the total intensity and the spectral index of the emission at the central frequency (parameter nterms=2 in the CASA task \emph{clean}). From this procedure, we obtain a model of the emission consisting of an elongated source at 32.5~GHz with a flux density of $\sim$175 $\mu$Jy, a spectral index of $\sim$0.7 and a deconvolved size of $\sim$22~m.a.s. with a P.A. of $\sim$35$^\circ$. In the context of Reynolds (1986), a spectral index of 0.7 implies an exponent of $-$0.6 in the power-law that describes the change of the size with frequency. In summary, our best model for the radio jet emission is given by

\begin{equation}
\left[ \frac{F_\nu}{\mu \rm{Jy~beam^{-1}}} \right] = 175  \times \left[ \frac{\nu}{32.5 \quad \rm{GHz}} \right]^{0.7}  \quad ; \quad \left[ \frac{\theta_\nu}{ \rm{m.a.s.}} \right] = 22 \times \left[ \frac{\nu}{32.5 \quad \rm{GHz}} \right]^{-0.6} 
\end{equation}

From this, we predict sizes of the jet of $\sim$27, $\sim$22, and $\sim$20~m.a.s. at K, Ka and Q bands, respectively. Note that these sizes imply that bright (optically thick) free-free emission is expected only at radii $<$1.5~au. In order to remove the free-free contamination in the VLA images of the dust emission, we subtracted the model components obtained from the MFS cleaning process to the uv data of the three VLA bands. 

\newpage

\section{Relationship between Model Parameters} \label{modvsobs}

\begin{figure}[h]
\begin{center}
\includegraphics[width=\textwidth]{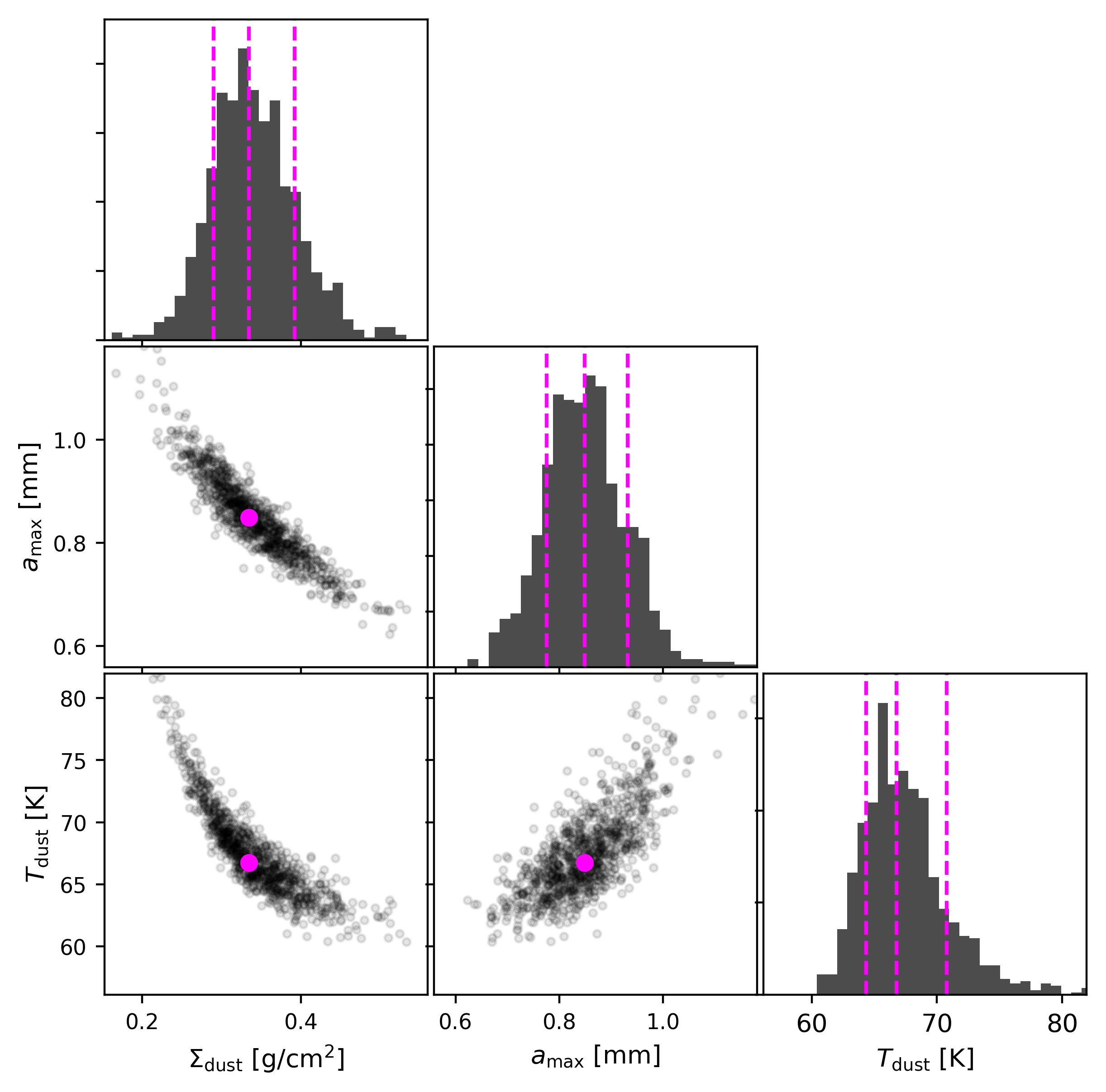}
\caption{Representation of the relation between the three parameters of the model: $T_{\rm dust}$, $\Sigma_{\rm dust}$, and $a_{\rm max}$. The top diagonal show the 1D histograms of the distribution of the fitted parameters. The vertical dashed lines represent the 16th, the 50th and the 84th percentiles. The 2D plots represent the bi-variate distributions for each pair of parameters. The magenta point in the 2D plots mark the best-fit value (50th percentile). The plot show the results obtained for the 1000 iterations at a intermediate radius (50~au).}
\label{FigB2}
\end{center}
\end{figure}

\newpage

\section{Comparison between Absorption-only approximation and Absorption+Scattering} \label{comparison}

\begin{figure}[h]
\begin{center}
\includegraphics[width=0.8\textwidth]{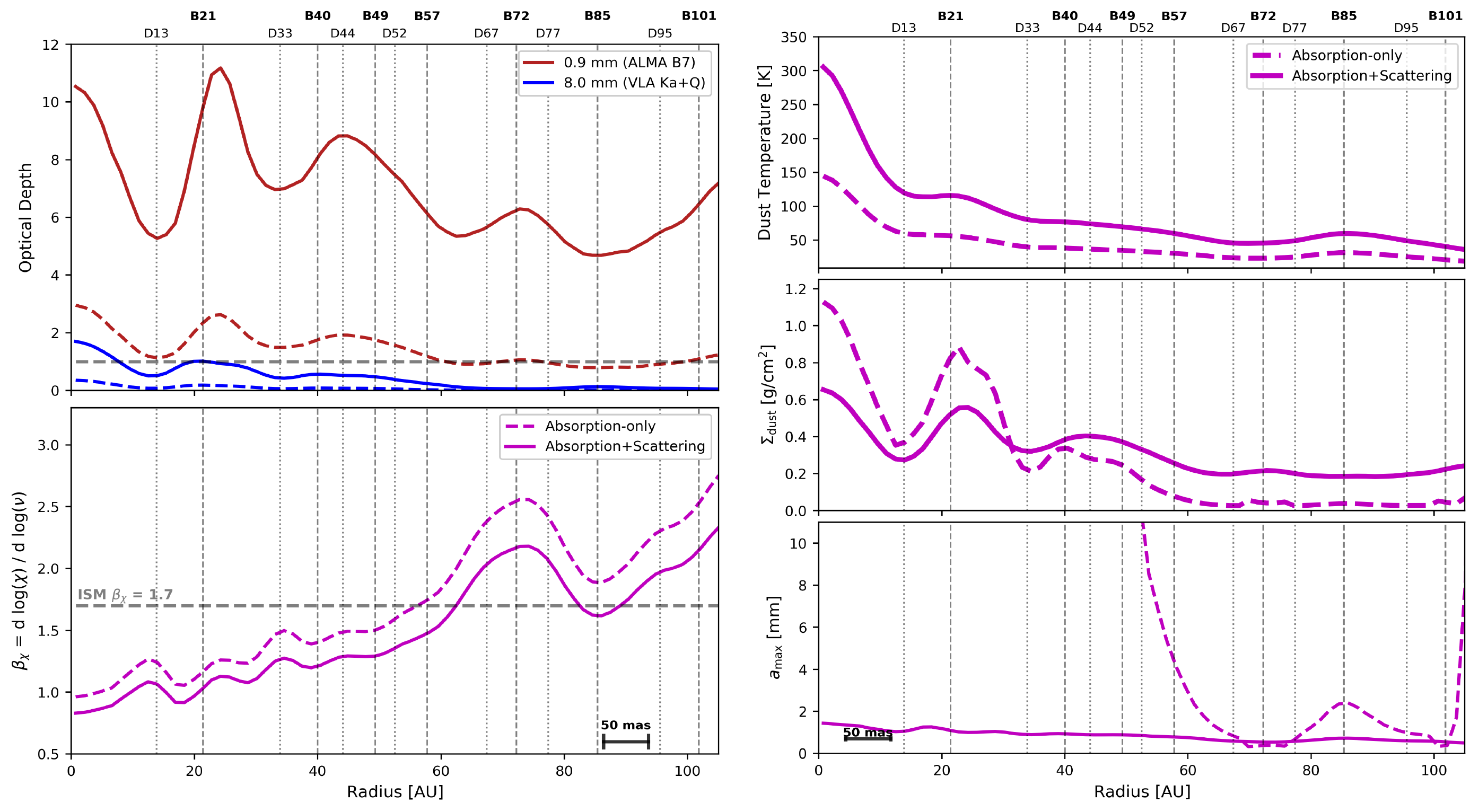}
\caption{Comparison between the results with the absorption-only approximation (dashed lines) and absorption+scattering (solid lines).}
\label{FigC1}
\end{center}
\end{figure}

 In Figure \ref{FigC1} we show a comparison of the results of the radial SED fitting obtained considering that opacity is dominated only by absorption (equation \ref{SEDabs}; dashed lines) and to the more complete equation considering both effects, absorption and scattering (equation \ref{SEDsca}; solid lines). The main errors that the absorption-only approximation introduces in the analysis can be summarized as follows:
 
\begin{itemize}

\item{} Optical depths are underestimated at all wavelengths. The value of $\beta$, which defines the variation of the optical depth with frequency, is overestimated. Note that for radii $<$60~au, the obtained values of $\beta$ are not consistent with the initial assumption of $\omega=0$ (see Figure \ref{Fig4}), i.e., the initial assumption of opacity dominated by absorption is not fulfilled for, at least, radii $\lesssim$60~au.

\item{} Dust temperature is underestimated, which has important implications in the positions of the snow lines.

\item{} Particles are estimated to be much larger at the internal parts of the disk. This is a direct consequence of comparing the obtained value of $\beta$ with that of the slope of the absorption coefficient, $\beta_\kappa$ which, for $a_{\rm max}\gtrsim$500~$\mu$m takes values much lower than $\beta_\chi$ (see Figure \ref{Fig4}). Then, in order to explain low values of $\beta$ only by absorption, we need to invoke the presence of much larger particles.

\item{} The low optical depths inferred and the lower values of the absorption coefficient with respect to the extinction coefficient, lead to very low values of the dust surface density at the external parts of the disk. At central parts of the disk, the extremely large sizes of the dust particles makes the disk much denser than it actually is. 

\item{} Since most of the mass is stored at large radii where low values of the density are obtained, the total dust mass of the disk is underestimated. In the case of HL Tau, we obtain a total dust mass of $\sim$0.5$\times$10$^{-3}$~$M_\odot$ in the absorption-only approximation. This is a $\sim$50\% lower than the value obtained considering absorption+scattering, $\sim$1.0$\times$10$^{-3}$~$M_\odot$.

\end{itemize}

\begin{figure}[h]
\begin{center}
\includegraphics[width=0.7\textwidth]{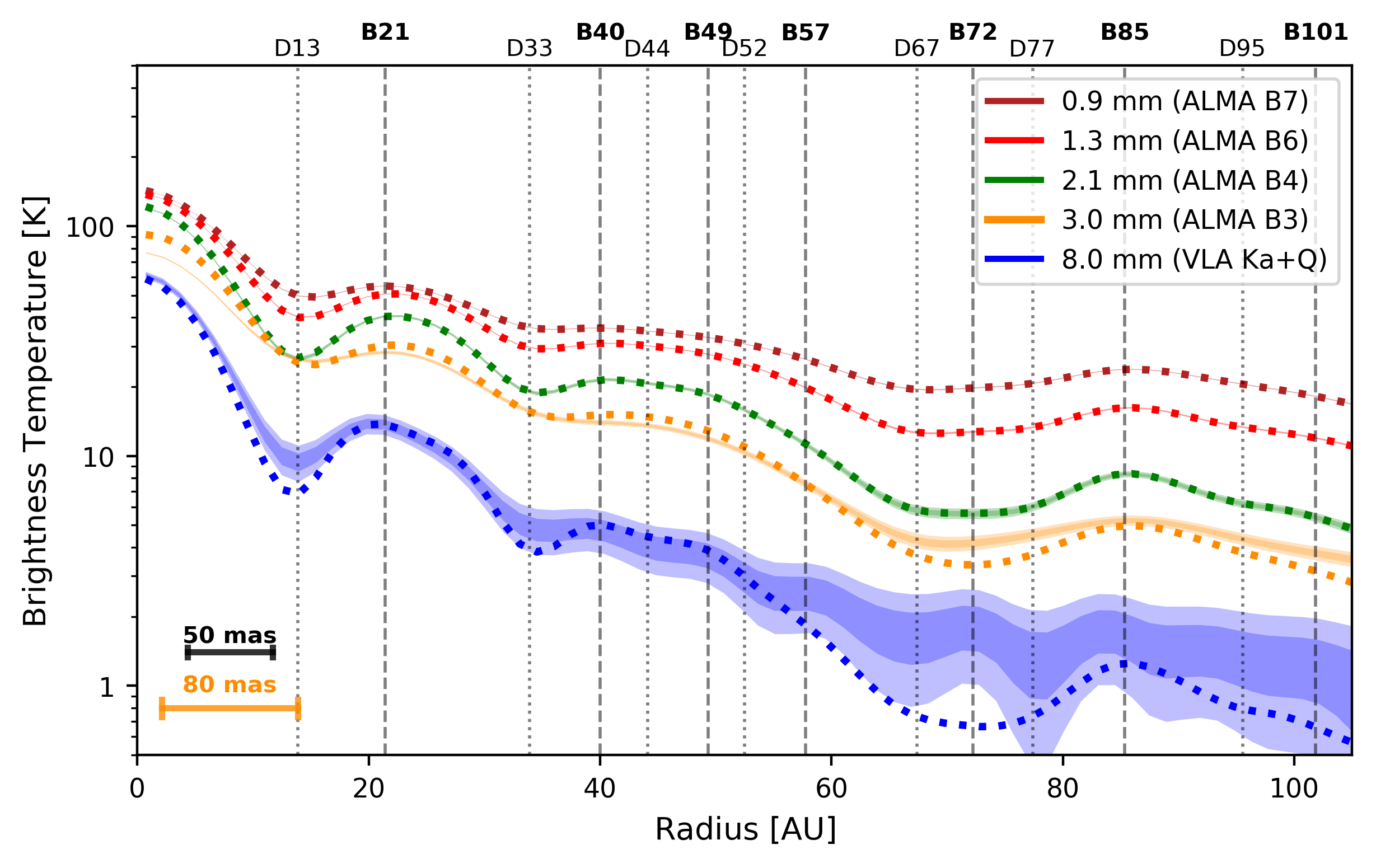}
\caption{Comparison between the predicted brightness temperatures at each wavelength from our absorption+only model, with the VLA and ALMA observations. Predictions from the model are shown as pointed lines. For the observations we show uncertainties at 1- and 2-$\sigma$ levels. For our analysis, we only take into account data with the highest angular resolution (50 m.a.s.; ALMA bands 7, 6, and 4 and VLA Ka+Q band). We also include a comparison of the prediction of our model with a lower angular resolution ($\sim$80 m.a.s.) ALMA data at 3.0 mm (band 3) after convolution of our results at this wavelength. While it is possible to find an absorption-only model with a good agreement with all observations, several errors are introduced in the obtained parameters by ignoring scattering effects (see Figure \ref{FigC1}).}
\label{FigC2}
\end{center}
\end{figure}

In Figure \ref{FigC2} we show a comparison between the absorption-only model and the observed data. While we can find a model with a good agreement with the observations, the wrong assumption of dust opacity dominated by absorption lead to the errors in the physical parameters described above. 

\newpage

\section{Flux Calibration Uncertainties} \label{FluxUnc}

 The analysis presented in this paper only takes into account statistical errors based on the rms noise of the images (see \S \ref{FinalImages}). The results could potentially be affected by additional uncertainties due to the flux calibration of each image. This uncertainty has a different nature and behaviour than statistical errors. For instance, flux calibration uncertainties will affect in the same way a whole image at a given wavelength. The effect is that the whole radial profile is moved to higher or lower values, but relative differences in the same image are conserved, i.e., local minima and local maxima of emission will appear always at the same positions. Thus, when flux calibration uncertainties are taken into account, radial profiles of each parameter obtained from these images might also increase or decrease their values, but always conserving the relative morphology. 
 
 According to the observatories, nominal flux uncertainties are $\sim$10\% or ALMA band 7 and VLA bands Q and Ka, and $\sim$5\% for ALMA bands 6 and 4. These values are for standard single epoch observations involving one observation of several minutes of a flux calibrator. In our case, observations at each band used several epochs: 10, 9, 4 epochs at bands 7, 6, 4, and Q+Ka, respectively. The final images are made by combining data of all epochs, thus, reducing the final flux uncertainty. While the true final uncertainty in flux density is hard to estimate, we adopt a conservative approximation in which we consider that it is by a factor of the square root of the number of epochs. This results in flux uncertainties of $\sim$3, 1.5, 2.5, and 3.5\% at bands 7, 6, 4, and Q+Ka, respectively. 

\begin{figure}[h]
\begin{center}
\includegraphics[width=0.8\textwidth]{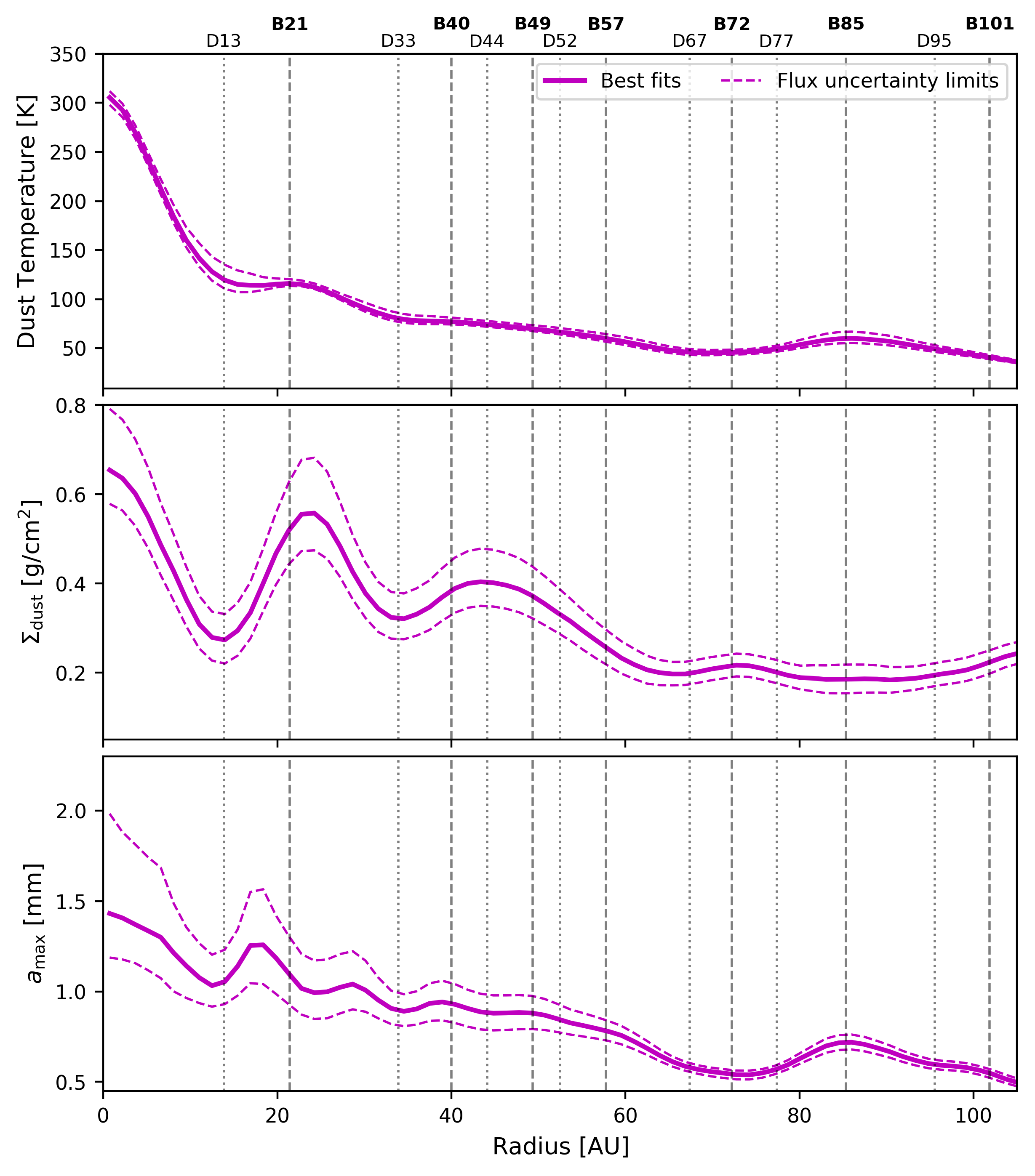}
\caption{The effect of the flux calibration uncertainty in the fitted parameters. Errors in the flux scale of the image due to flux calibration affect in the same way to the whole image at a given wavelength. When taken into account in the radial fit of the SED, the effect is that the radial profiles of each parameter can be moved to higher or lower values, but always conserving the shape of the profile. This is, local maxima and minima of each parameter will appear always at the same positions. The limits within each parameter can vary are marked in the figure as dashed lines. The effect of the flux scale uncertainty dominates at the central parts of the disk, where emission is detected with higher signal-to-noise, while at radii $>$50 au, statistical errors based on the rms of the maps dominate.}
\label{FigD1}
\end{center}
\end{figure}

 To explore the effect of the flux calibration uncertainty in the derived parameters, we repeated the procedure explained in \S \ref{SEDfit} 1000 times, but each time we randomly vary each of the radial profiles at different wavelengths within their flux uncertainties following a normal distribution. As expected, these uncertainties affect mainly at the center of the disk, where higher signal-to-noise are present, and then, where the values of the flux can vary in a wider range. It is important to emphasize that the result of this is to move the whole resulting radial profiles of each parameter to higher or lower values, but always conserving their morphology. This is, the values of each parameter at different radii cannot independently take any value; local minima and local maxima should be always located at the same positions. The result of this is shown in Figure \ref{FigD1}, where we show as dashed lines the ranges within each radial profile can vary. The lower limit is defined by the percentile 16 and the upper limit as the percentile 84 of our 1000 fits. 

 In our case, the effect of the flux calibration uncertainties is to slightly increase or decrease the dust surface density and the maximum particle size at the central parts of the disk ($<$60~au). The value of the maximum particle size at the center of the disk can vary in the range $\sim$1-2~mm. There is not great impact on the dust temperature, which leave the snow lines at the same positions than discussed in section \S \ref{nature}. We emphasized that the flux scale uncertainty does not change the positions of the local maxima and minima of the parameters. Thus, we still obtain a good agreement between the positions of the snow lines and the accumulations of large or small particles.

\end{document}